\RequirePackage[displaymath]{lineno}
\documentclass[twocolumn,aps,prc,showpacs,superscriptaddress,floatfix]{revtex4-2}
\usepackage{newtxtext,newtxmath,booktabs,siunitx}
\usepackage{dcolumn}
\usepackage{bm}
\usepackage{float}
\usepackage[]{graphicx}  
\usepackage{booktabs}
\usepackage{tabularx}
\usepackage{amsmath,bm}
\usepackage{xcolor}
\usepackage{multirow}
\usepackage[colorlinks,citecolor=blue,urlcolor=blue,linkcolor=blue]{hyperref}
\usepackage[displaymath]{lineno}

\usepackage{rotating}
\usepackage{epsfig} 
\usepackage{graphicx}
\usepackage[caption=false]{subfig}

\graphicspath{ {./}, {./plots_pdf/} }
\DeclareGraphicsExtensions{ .pdf, .png}

\usepackage{xcolor}
\usepackage{xparse,xcoffins}

\ExplSyntaxOn
\NewCoffin\imagecoffin
\NewCoffin\labelcoffin

\keys_define:nn { fqwang/label }
 {
  label   .tl_set:N = \l_fqwang_label_tl,
  labelbox .bool_set:N = \l_fqwang_label_box_bool,
  labelbox .default:n = true,
  fontsize .tl_set:N = \l_fqwang_label_size_tl,
  fontsize .initial:n = \normalsize,
  pos .choice:,
  pos/nw .code:n = \tl_set:Nn \l_fqwang_label_pos_tl { left,up },
  pos/ne .code:n = \tl_set:Nn \l_fqwang_label_pos_tl { right,up },
  pos/sw .code:n = \tl_set:Nn \l_fqwang_label_pos_tl { left,down },
  pos/se .code:n = \tl_set:Nn \l_fqwang_label_pos_tl { right,down },
  pos/nwhigh .code:n = \tl_set:Nn \l_fqwang_label_pos_tl { left,uphigh },
  pos/nehigh .code:n = \tl_set:Nn \l_fqwang_label_pos_tl { right,uphigh },
  pos/swhigh .code:n = \tl_set:Nn \l_fqwang_label_pos_tl { left,downhigh },
  pos/sehigh .code:n = \tl_set:Nn \l_fqwang_label_pos_tl { right,downhigh },
  pos/nwlow .code:n = \tl_set:Nn \l_fqwang_label_pos_tl { left,uplow },
  pos/nelow .code:n = \tl_set:Nn \l_fqwang_label_pos_tl { right,uplow },
  pos/swlow .code:n = \tl_set:Nn \l_fqwang_label_pos_tl { left,downlow },
  pos/selow .code:n = \tl_set:Nn \l_fqwang_label_pos_tl { right,downlow },
  pos/nwHigh .code:n = \tl_set:Nn \l_fqwang_label_pos_tl { left,upHigh },
  pos/neHigh .code:n = \tl_set:Nn \l_fqwang_label_pos_tl { right,upHigh },
  pos/swHigh .code:n = \tl_set:Nn \l_fqwang_label_pos_tl { left,downHigh },
  pos/seHigh .code:n = \tl_set:Nn \l_fqwang_label_pos_tl { right,downHigh },
  pos/nwLow .code:n = \tl_set:Nn \l_fqwang_label_pos_tl { left,upLow },
  pos/neLow .code:n = \tl_set:Nn \l_fqwang_label_pos_tl { right,upLow },
  pos/swLow .code:n = \tl_set:Nn \l_fqwang_label_pos_tl { left,downLow },
  pos/seLow .code:n = \tl_set:Nn \l_fqwang_label_pos_tl { right,downLow },
  pos/n .code:n = \tl_set:Nn \l_fqwang_label_pos_tl { hc,up },
  pos/w .code:n = \tl_set:Nn \l_fqwang_label_pos_tl { left,vc },
  pos/s .code:n = \tl_set:Nn \l_fqwang_label_pos_tl { hc,down },
  pos/e .code:n = \tl_set:Nn \l_fqwang_label_pos_tl { right,vc },
  pos .initial:n = nw,
  unknown .code:n   = \clist_put_right:Nx \l_fqwang_label_clist
                       { \l_keys_key_tl = \exp_not:n { #1 } }
 }

\clist_new:N \l_fqwang_label_clist
\box_new:N \l_fqwang_label_box
\box_new:N \l_fqwang_label_image_box

\NewDocumentCommand{\xincludegraphics}{O{}m}
 {
  \group_begin:
  \tl_clear:N \l_fqwang_label_tl
  \clist_clear:N \l_fqwang_label_clist
  \keys_set:nn { fqwang/label } { #1 }
  \tl_if_empty:NTF \l_fqwang_label_tl
   {
    \fqwang_includegraphics:Vn \l_fqwang_label_clist { #2 }
   }
   {
    \SetHorizontalCoffin\imagecoffin
     {
      \fqwang_includegraphics:Vn \l_fqwang_label_clist { #2 }
     }
    \SetHorizontalCoffin\labelcoffin
     {
      \raisebox{\depth}
       {
        \bool_if:NTF \l_fqwang_label_box_bool
         { \fcolorbox{white}{white}{\l_fqwang_label_size_tl\l_fqwang_label_tl} }
         { \l_fqwang_label_size_tl\l_fqwang_label_tl }
       }
     }
    \SetVerticalPole\imagecoffin{left}{36pt+\CoffinWidth\labelcoffin/2}
    \SetVerticalPole\imagecoffin{right}{\Width-36pt-\CoffinWidth\labelcoffin/2}
    \SetHorizontalPole\imagecoffin{up}{\Height-12pt-\CoffinHeight\labelcoffin/2}
    \SetHorizontalPole\imagecoffin{down}{12pt+\CoffinHeight\labelcoffin/2}
    \SetHorizontalPole\imagecoffin{uphigh}{\Height-6pt-\CoffinHeight\labelcoffin/2}
    \SetHorizontalPole\imagecoffin{downlow}{6pt+\CoffinHeight\labelcoffin/2}
    \SetHorizontalPole\imagecoffin{uplow}{\Height-18pt-\CoffinHeight\labelcoffin/2}
    \SetHorizontalPole\imagecoffin{downhigh}{18pt+\CoffinHeight\labelcoffin/2}
    \SetHorizontalPole\imagecoffin{upHigh}{\Height-0pt-\CoffinHeight\labelcoffin/2}
    \SetHorizontalPole\imagecoffin{downLow}{0pt+\CoffinHeight\labelcoffin/2}
    \SetHorizontalPole\imagecoffin{upLow}{\Height-24pt-\CoffinHeight\labelcoffin/2}
    \SetHorizontalPole\imagecoffin{downHigh}{24pt+\CoffinHeight\labelcoffin/2}
    \use:x{\JoinCoffins\imagecoffin[\l_fqwang_label_pos_tl]\labelcoffin[vc,hc]} 
    \TypesetCoffin\imagecoffin
   }
   \group_end:
 }
\NewDocumentCommand{\setlabel}{m}
 {
  \keys_set:nn { fqwang/label } { #1 }
 }
\cs_new_protected:Nn \fqwang_includegraphics:nn
 {
  \includegraphics[#1]{#2}
 }
\cs_generate_variant:Nn \fqwang_includegraphics:nn { V }

\ExplSyntaxOff

\begin{document}
\newcommand{\snn}       {\sqrt{s_{_{\rm NN}}}}
\newcommand{\pp}       {$p+p$}
\newcommand{\pa}       {$p+A$}
\newcommand{\pythia}   {Pythia}
\newcommand{\pmodel}   {Pythia 8.303}
\newcommand{\Nch}      {N_{\rm ch}}
\newcommand{\Ntot}     {N_{\rm tot}}
\newcommand{\pt}       {p_\perp}
\newcommand{\gevc}     {GeV/$c$}
\newcommand{\dphi}     {\Delta\phi}
\newcommand{\deta}     {\Delta\eta}
\newcommand{\Vlarge}   {V_n^{\mbox{{\footnotesize large}-}\deta}}
\newcommand{\Vsub}     {V_n^{\rm sub}}
\newcommand{\mean}[1]  {\langle #1 \rangle}

\newcommand{\red}[1]   {\textcolor{red}{#1}}
\newcommand{\blue}[1]  {\textcolor{blue}{#1}}
\newcommand{\green}[1] {\textcolor{green}{#1}}
\newcommand{\cyan}[1]  {\textcolor{cyan}{#1}}
\newcommand{\rep}[2]   {\gray{#1}\green{#2}}

\title{Azimuthal Correlation Anisotropies in $p+p$ Collisions Simulated using  \pythia}

\author{Manuel Sebastian Torres}
\affiliation{Department of Physics, Universidad Nacional de Colombia,  Bogotá D.C. 111321, Colombia }
\affiliation{Department of Physics and Astronomy, Purdue University, West Lafayette, Indiana 47907, USA}
\author{Yicheng Feng}
\affiliation{Department of Physics and Astronomy, Purdue University, West Lafayette, Indiana 47907, USA}
\author{Fuqiang Wang}
\affiliation{Department of Physics and Astronomy, Purdue University, West Lafayette, Indiana 47907, USA}

\date{\today}
\begin{abstract}
Stimulated by the keen interest in possible collective behavior in high-energy proton-proton and proton-nucleus collisions, we study two-particle angular correlations in pseudorapidity and azimuthal differences in simulated \pp\ interactions using the Pythia 8 event generator. Multi-parton interactions and color connection are included in these simulations which have been perceived to produce collectivity in final-state particles. Meanwhile, contributions from genuine few-body nonflow correlations, not of collective flow behavior, are known to be severe in these small-system collisions. We present our Pythia correlation studies pedagogically and report azimuthal harmonic anisotropies analyzed using several methods. We observe anisotropies in these Pythia simulated events qualitatively and semi-quantitatively similar to experimental data. Our findings highlight the delicate nature of azimuthal anisotropies in small-system collisions, and provide a benchmark that can aid in improving data analysis and interpreting experimental measurements in small-system collisions.
\end{abstract}

\maketitle


  
\section{Introduction}

One of the most interesting aspects of heavy ion collisions is the azimuthal angle distribution of particles in the final state. This is because the overlap geometry of the collision is anisotropic, like an almond shape. This anisotropic geometry is imprinted on the final state particle momentum distribution because of interactions among the particles before they freeze out~\cite{Ollitrault:1992bk}. These interactions drive the collective expansion of the overlap collision system, a phenomenon called collective flow~\cite{Heinz:2013th}. The comparison between the momentum anisotropy and the initial geometry anisotropy can provide information about the interactions in the hot and dense medium, the quark-gluon plasma (QGP), created in those collisions. 

To study azimuthal anisotropies, particle distributions are often written in Fourier harmonic series in terms of the particle azimuthal angle ($\phi$) relative to the reaction plane ($\Psi$)~\cite{Voloshin:1994mz}:
\begin{equation}
    \frac{dN}{d\phi} \propto 1 + \sum_{n=1}^{\infty}2v_n\cos n(\phi-\Psi)\,.
    \label{eq:dNdphi}
\end{equation}
The coefficients, $v_n$, characterize the harmonic anisotropies.
The most straightforward way to calculate the anisotropies is through two-particle  correlations~\cite{Poskanzer:1998yz}:
\begin{equation}
    \frac{dN}{d\dphi} \propto 1 + \sum_{n=1}^{\infty}2V_n\cos n\dphi\,,
\end{equation}
where $\dphi=\phi_1-\phi_2$ is the pair azimuthal opening angle. The Fourier coefficients, characterizing the two-particle cumulants, are given by
\begin{equation}
    V_n = \langle \cos n\dphi \rangle\,.
	\label{eq:Fourier}
\end{equation}
If the azimuthal anisotropy in $\dphi$ comes all from global correlations of all particles with respect to the reaction plane (i.e.~flow), then $v_n = \sqrt{V_n}$. 

However, two-particle correlations can result  from processes like resonance decays and jets. These few-body correlations contribute to $V_n$, and are often referred to as nonflow~\cite{Borghini:2000cm,Wang:2008gp,Ollitrault:2009ie}, in contrast to the global flow  correlations. Thus, $V_n$ contains both flow and nonflow contributions.
Because nonflow arises from two- and few-body particle correlations, their effect is diluted by particle multiplicity, resulting in a typical inverse multiplicity dependence in the two-particle cumulants,
\begin{equation}
    V_n = v_n^2 + C/N\,.
    \label{eq:Vn}
\end{equation}
Because of the large multiplicity in mid-central and central heavy-ion collisions, nonflow contamination is minor compared to collective flow~\cite{Feng:2024eos}. Nonflow can dominate in peripheral collisions where multiplicity is small.

In heavy ion collision studies, proton-proton (\pp) and proton-nucleus (\pa) collisions are often taken as references to infer properties of the QGP, or nuclear effects in hot QCD (quantum chromodynamics). 
It has long been considered that the particle azimuthal distributions in those small-system (\pp\ and \pa) collisions should be uniform (no preferred azimuthal direction exists), and the two-particle anisotropy $V_n$ in those small-system collisions must all be nonflow.
However, recent experiments suggest that collective flow may exist in \pa\ and \pp\ collisions, particularly in high-multiplicity events~\cite{Khachatryan:2010gv,CMS:2012qk,Aad:2012gla,Abelev:2012ola,ABELEV:2013wsa,Adamczyk:2015xjc,ATLAS:2016yzd,PHENIX:2017xrm,PHENIX:2018lia,PHENIX:2021ubk,STAR:2022pfn,ALICE:2023ulm,STAR:2023wmd}; one strong indication is that the four- and multi-particle cumulants, essentially devoid of nonflow, are finite~\cite{CMS:2013jlh,Aad:2013fja,ALICE:2019zfl}.
In two-particle cumulant measurements of Eq.~(\ref{eq:Vn}), the nonflow effect can be severe because the event multiplicity, and thus the multiplicity dilution, is small. This is the case even in high-multiplicity events of those small-system collisions, where the multiplicity is comparable to that in peripheral heavy ion collisions.
Many nonflow mitigation methods have been used~\cite{Li:2012hc,Nagle:2018nvi,Feng:2024eos}.
After nonflow mitigation, it has been found in small system collisions, especially at LHC energies, that finite positive $v_n^2$ remain~\cite{Nagle:2018nvi}. This led to the conclusion that a small QGP droplet may be formed in small system collisions~\cite{Nagle:2018nvi}. 

On the other hand, the overlap geometry in small-system collisions is lumpy due to fluctuations, and final-state momentum anisotropy could result if there is some degree of interactions~\cite{He:2015hfa,Romatschke:2017vte,Kurkela:2018oqw,Kurkela:2021ctp}, e.g., multi-parton interactions~\cite{dEnterria:2010xip,Ortiz:2013yxa,Bierlich:2017vhg,Sjostrand:2018xcd}, in these collisions. 
The magnitude of those interactions, gauged by the magnitude of the final-state momentum anisotropy relative to the initial geometry anisotropy,  would then discriminate whether the interactions occur in a low-density environment or a high-density QGP.

In addition to these physics questions, the technical question of how well nonflow has been mitigated is not settled and is under extensive investigation. 
It is therefore interesting to examine anisotropies in model simulations of \pp\ collisions, particularly using the  \pythia\ event generator~\cite{Sjostrand:1993yb,Sjostrand:2014zea,Sjostrand:2019zhc,Bierlich:2022pfr}. Pythia is  naively not expected to generate single-particle azimuthal anisotropy because it is not a cascade model with final-state interactions. 
However, recent implementation of multi-parton interactions (color reconnection) is perceived to mimic some degree of final-state interactions; thus we cannot completely exclude collective flow~\cite{Sjostrand:2014zea,Sjostrand:2017cdm,Bierlich:2018xfw}.
Studies of anisotropies in  \pythia\ would, therefore, be valuable in at least two fronts. One is to aid in studying nonflow subtraction techniques and associated uncertainties~\cite{Feng:2024eos,Lim:2019cys,Abdulhamid:2024ifx}, and the other is to discern any collective flow in \pp\ interactions to guide the interpretations of  experimental anisotropy measurements in small-system collisions~\cite{Li:2012hc,Nagle:2018nvi}.  

We hereby report a study on azimuthal correlation anisotropies in  \pp\ interactions simulated using  Pythia  version 8.303~\cite{Bierlich:2022pfr}, which includes multi-parton interactions and color reconnection~\cite{Sjostrand:2017cdm,Bierlich:2018xfw}. 

\section{Pythia Model Setup}
We use  \pmodel~\cite{Bierlich:2022pfr} to simulate \pp\ collisions at the center-of-mass energy of $\snn=200$~GeV. We generate $3\times10^8$ events each for diffractive and non-diffractive interactions with soft processes (SoftQCD)~\cite{Sjostrand:2014zea}. 
The soft processes in  \pythia\ represent the entire cross-section ($\sigma_{\rm tot}$) of the hadron-hadron interaction, i.e.~minimum bias (MB) interaction. 
Rare processes, like jet physics at all scales, occur as part of the total cross-section.
In the case of \pp\ collision in  \pythia, the cross section is calculated using the 1992 Donnachie-Landshoff parameterization~\cite{Sjostrand:2014zea}. This total cross-section is the sum of several components:
$\sigma_{\rm tot} =  \sigma_{{\rm sd}} + \sigma_{\rm dd} + \sigma_{\rm nd} + \sigma_{\rm el}$~\cite{Sjostrand:2006za}. 
The diffractive part is divided into the single (sd) and the double (dd) component, where the former has the $pp \rightarrow  pX$ processes whereas the latter has the $pp \rightarrow ppX$ processes.
The diffractive events in our simulation include both single-diffractive and double-diffractive  processes. 
The non-diffractive (nd) part includes all other inelastic processes $pp \rightarrow X$. 
The elastic (el) process, $pp \rightarrow pp$, does not produce particles and thus not included in our simulation.  
The comparison between diffractive and non-diffractive interactions is expected to provide insight into initial- vs.~final-state effects, because diffractive interactions should be more closely related to initial state of the proton whereas non-diffractive interactions could be influenced by final-state effects if any.

Table~\ref{tab:list} lists the majority of particle species and their average compositions in non-diffractive events. 
In our simulations, we keep the electromagnetic- and weak-decay hadrons stable. 

\begin{table}[H]
\caption{Average full-phase space multiplicities of various particle species in non-diffractive events generated using \pmodel. A total of approximately $3 \times 10^8$ events are generated.} 
\label{tab:list}
\begin{center}
\begin{tabular}{cc|cc}
\hline
particle  & multiplicity & particle         & multiplicity  \\ \hline
$\pi^+$   & 12.140       & $\pi^-$          & 11.422        \\ 
$\pi^0$   & 13.289       &                  &               \\
$K^+$     & 1.631        & $K^-$            & 1.459         \\ 
$K^0$     & 2.263        & $\bar{K}^0$      & 1.059         \\
$p$         & 4.162        & $\bar{p}$        & 0.700       \\ 
$n$         & 1.768        & $\bar{n}$        &  0.699      \\ 
$\Lambda$ & 0.370        & $\bar{\Lambda}$  & 0.217         \\ \hline
\end{tabular}
\end{center}
\end{table}

\begin{figure}[H]
	\centering
    \xincludegraphics[width=0.45\textwidth,pos=nwlow,label=\hspace*{0.7cm}a)]{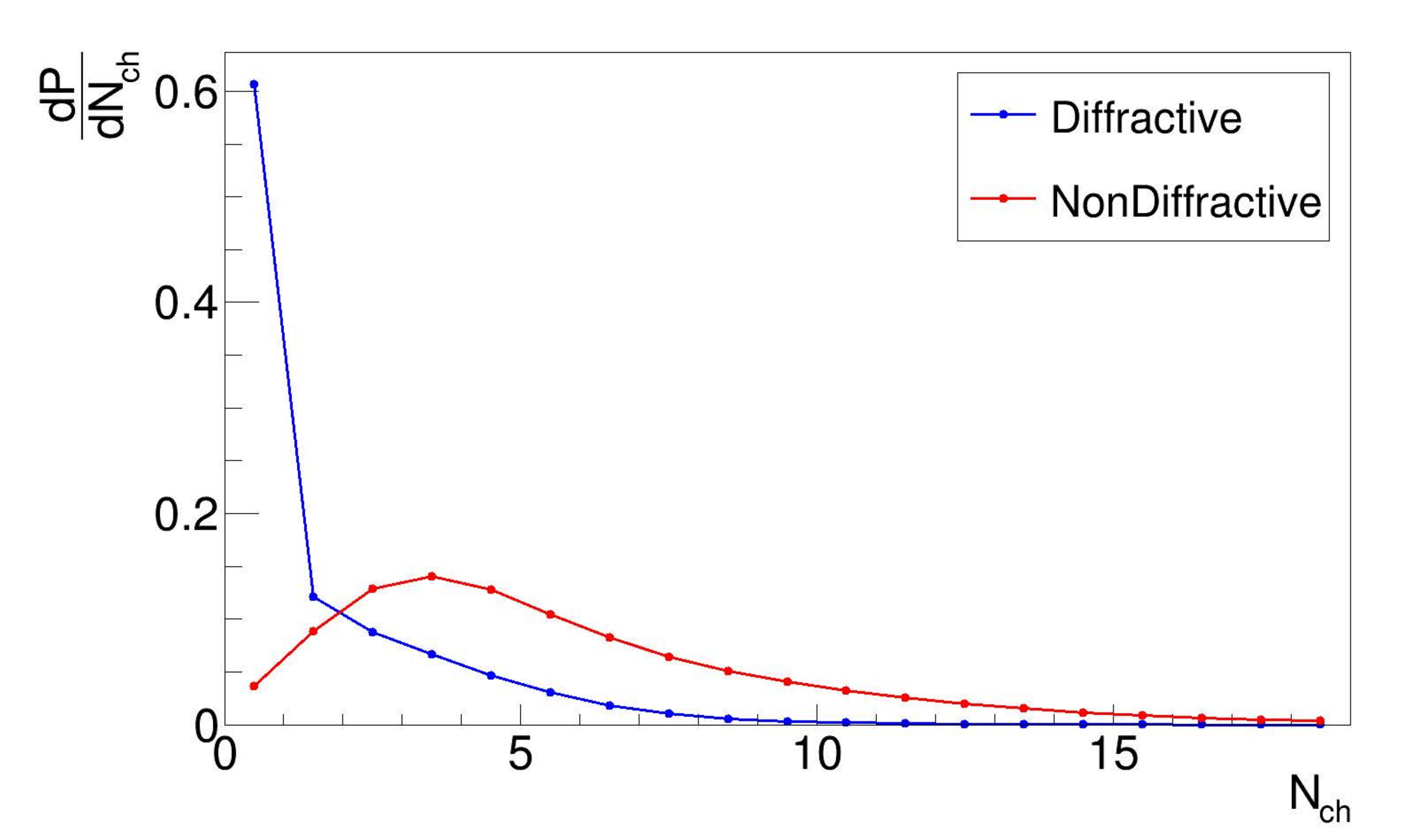}\hfill
	\xincludegraphics[width=0.45\textwidth,pos=nwlow,label=\hspace*{0.7cm}b)]{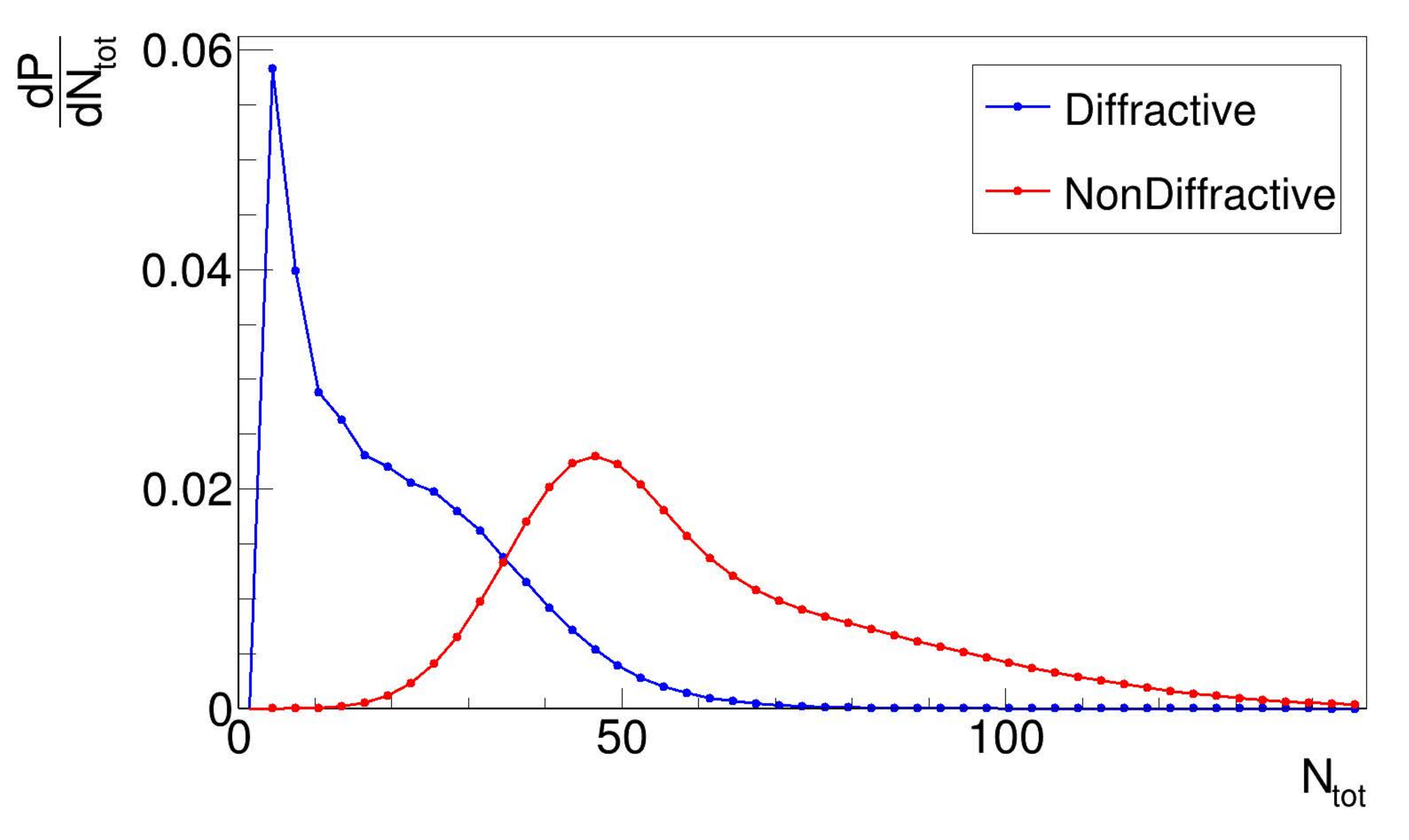}\hfill
    \caption{Probability distributions of diffractive and non-diffractive $p+p$ collisions simulated using \pmodel\ as a function of (a) $\Nch$, the midrapidity (within $|\eta|<1$ and $0.2<\pt<2$~\gevc) charged hadron ($\pi^\pm,K^\pm,p,\bar{p}$) multiplicity, and (b) $\Ntot$, the total final-state particle multiplicity (including all charged and neutral particles, over $4\pi$ solid angle).} 
    \label{fig:mult}
\end{figure}
We include only $\pi^\pm, K^\pm, p, \bar{p}$ in our study, referred to as particles of interest (POI).
Because experiments (such as STAR~\cite{Ackermann:2002ad} at RHIC and ALICE~\cite{Alme:2010ke} at the LHC) typically measure charged particles in the limited midrapidity region, we  focus on charged hadrons within the pseudorapidity range $|\eta|<1$ and transverse momentum range $0.2<\pt<2$~\gevc.  
Figure~\ref{fig:mult}(a) shows the diffractive and non-diffractive event probability distributions in POI multiplicity within $|\eta|<1$ and $0.2<\pt<2$~\gevc, referred to as $\Nch$ hereinafter. 
As reference the distribution in the total number of final-state particles ($\Ntot$), including both charged and neutral particles over all phase space, is shown in Fig.~\ref{fig:mult}(b).
Diffractive events have fewer produced particles than non-diffractive events, as expected. 

Events of differing $\Nch$ are expected to have varying physics, particularly in non-diffractive events, influences of final-state effects such as multi-parton interactions and color reconnection on possibly collective behavior.
We thus divide the events into the following $\Nch$ bins: 1--3, 4--6, 7--10, 11--14, 15--20.
We study anisotropies of POI in $\Nch$ bins and present our results as a function of the mean multiplicity $\mean{\Nch}$ of each $\Nch$ bin. 

\section{Correlation Analysis}

\subsection{Two-particle cumulant anisotropy}

\begin{figure}[H]
	\centering
	\includegraphics[width=0.48\textwidth]{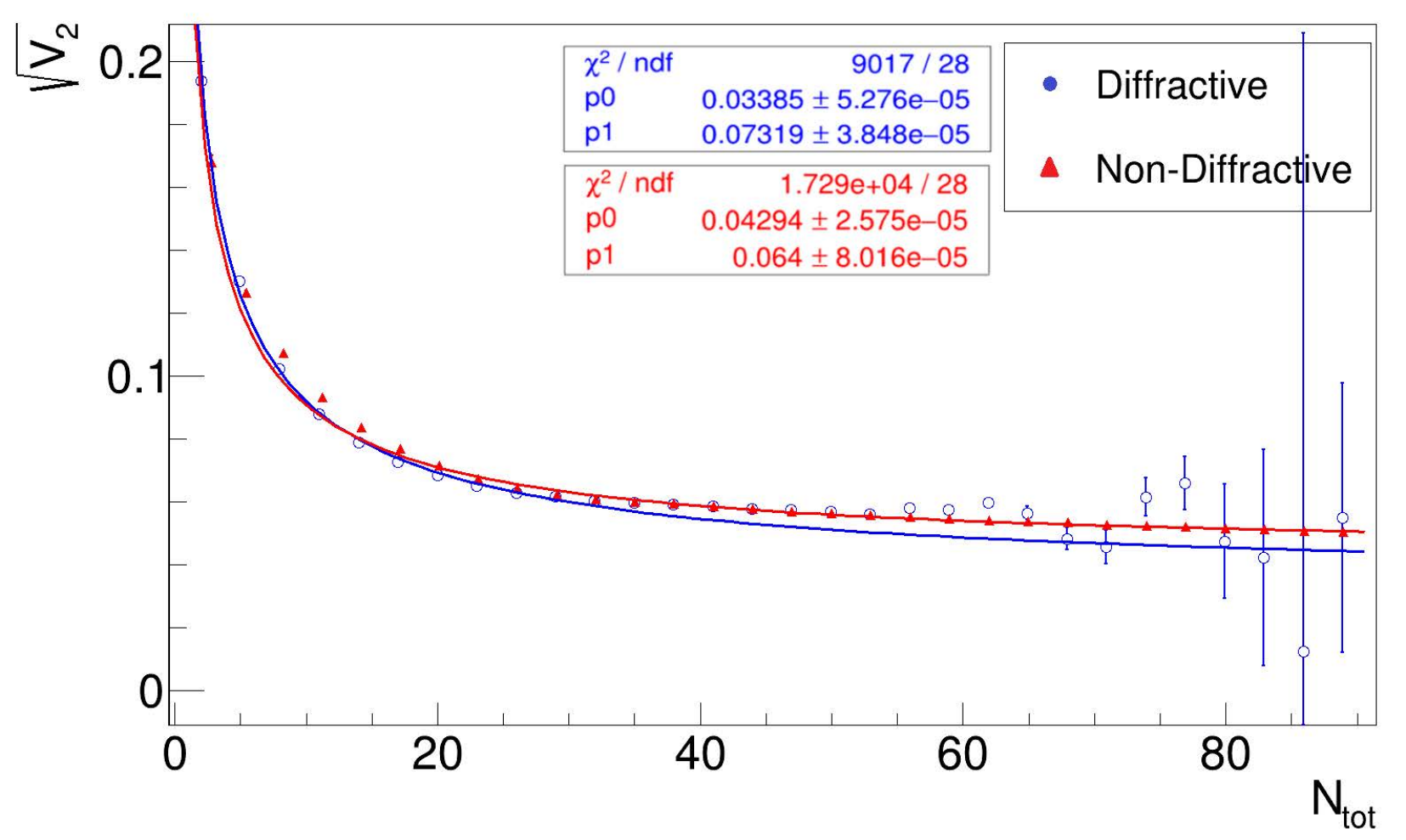} 
	\caption{Two-particle cumulant parameter, $\sqrt{V_2}$, as a function of the total multiplicity $N_{\rm tot}$. Fit curves are $\sqrt{V_2}=\sqrt{p_0^2+p_1/N_{\rm tot}}$ where $p_0$ and $p_1$ are fit parameters.}
	\label{fig:v2star}
\end{figure}

Figure~\ref{fig:v2star} shows $\sqrt{V_n}$ as a function of the total multiplicity $\Ntot$ for both diffractive and non-diffractive events.
It is observed that $\sqrt{V_2}$ decreases with $\Ntot$ and appears to saturate at large $\Ntot$. Motivated by Eq.~(\ref{eq:Vn}), we fit the data by $\sqrt{p_0^2+p_1/\Ntot}$. The fitted $p_0$ parameter is non-zero for both diffractive and non-diffractive events, suggesting that both types of events may have flow-like global anisotropies. The magnitudes of the $p_0$ parameter are comparable between diffractive and non-diffractive events.
This is rather surprising given the vastly different physics processes in diffractive and non-diffractive interactions.

Note that the fit quality is poor, indicating that the simple relation in Eq.~(\ref{eq:Vn}) may not hold. 
For example, the nonflow correlation strength such as jet correlations may increase faster than linearly in multiplicity which would result in a slower decrease of $V_n$ than $1/\Ntot$. The global flow-like correlation of the underlying event may also change with the event multiplicity so a constant $p_0$ parameter may not be adequate.
It is clear that a more rigorous investigation is required to identify the structure of \pp\ events from  \pythia\ simulations.

\subsection{Two-particle $(\deta,\dphi)$ correlations}
Two-particle $(\deta,\dphi)$ correlations are widely used to identify sources of correlations, where $\deta\equiv\eta_1-\eta_2$ and $\dphi\equiv\phi_1-\phi_2$ are the two-particle distances in pseudorapidity and azimuthal angle~\cite{Wang:2013qca}. For example, resonance decays manifest as a near-side peak at $(\deta,\dphi)\approx(0,0)$; so do intra-jet correlations. The effects of global momentum conservation and dijet correlations manifest as away-side ridge peaked at $\dphi=\pi$ and roughly uniform in $\deta$. 
Global momentum conservation may be modeled by a negative $V_1=\mean{\cos\dphi}$ term~\cite{Borghini:2006yk}, and others are often modeled by Gaussians. In our simulation of soft QCD interactions (i.e.~MB events), jet contributions are minimal. However, high-multiplicity events are known to be biased towards stronger jet production~\cite{Jacobs:2004qv}.

\begin{figure}[H]
	\centering
    \xincludegraphics[width=0.48\textwidth,labelbox,label=a)]{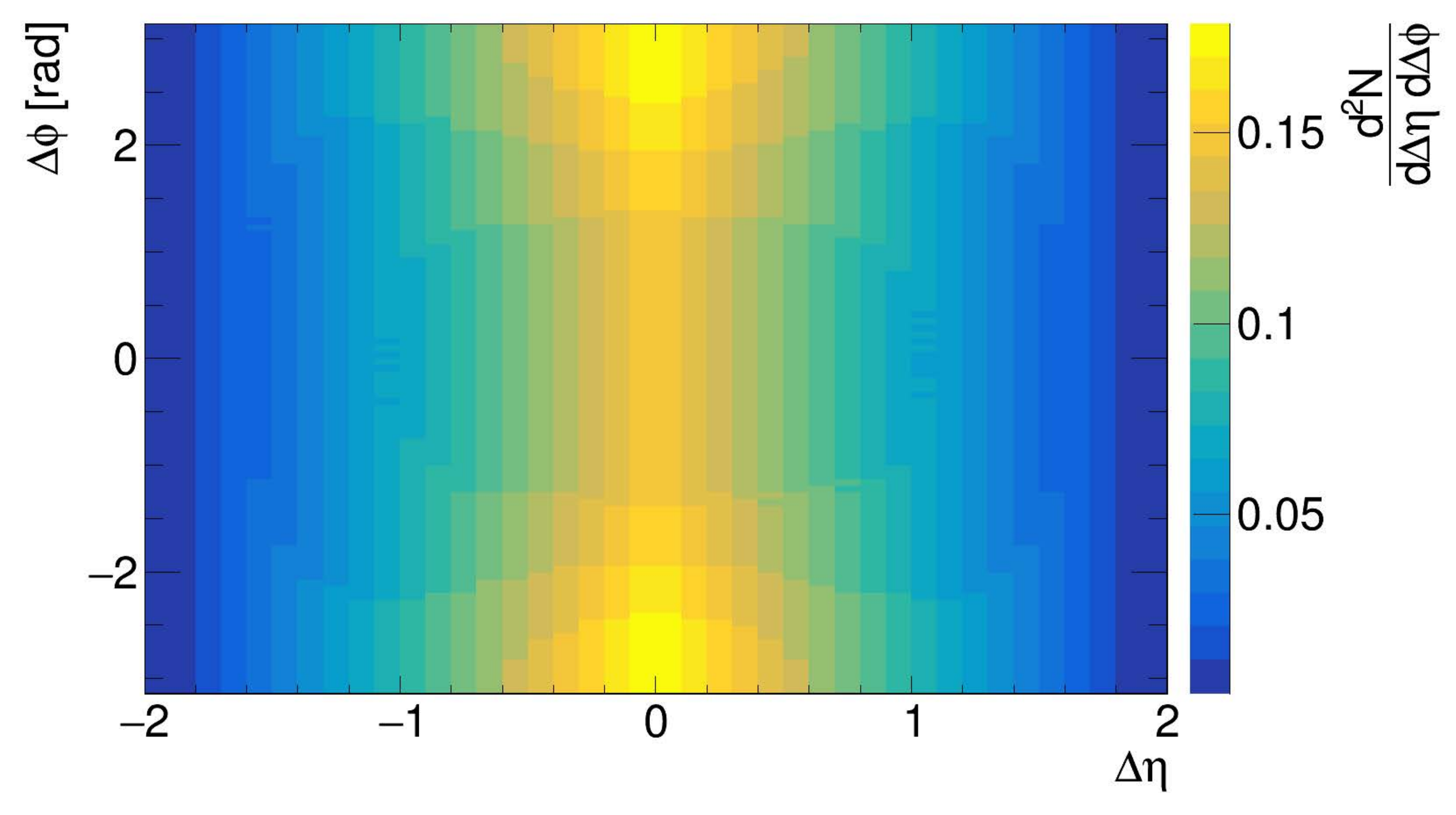}\hfill
	\xincludegraphics[width=0.48\textwidth,labelbox,label=b)]{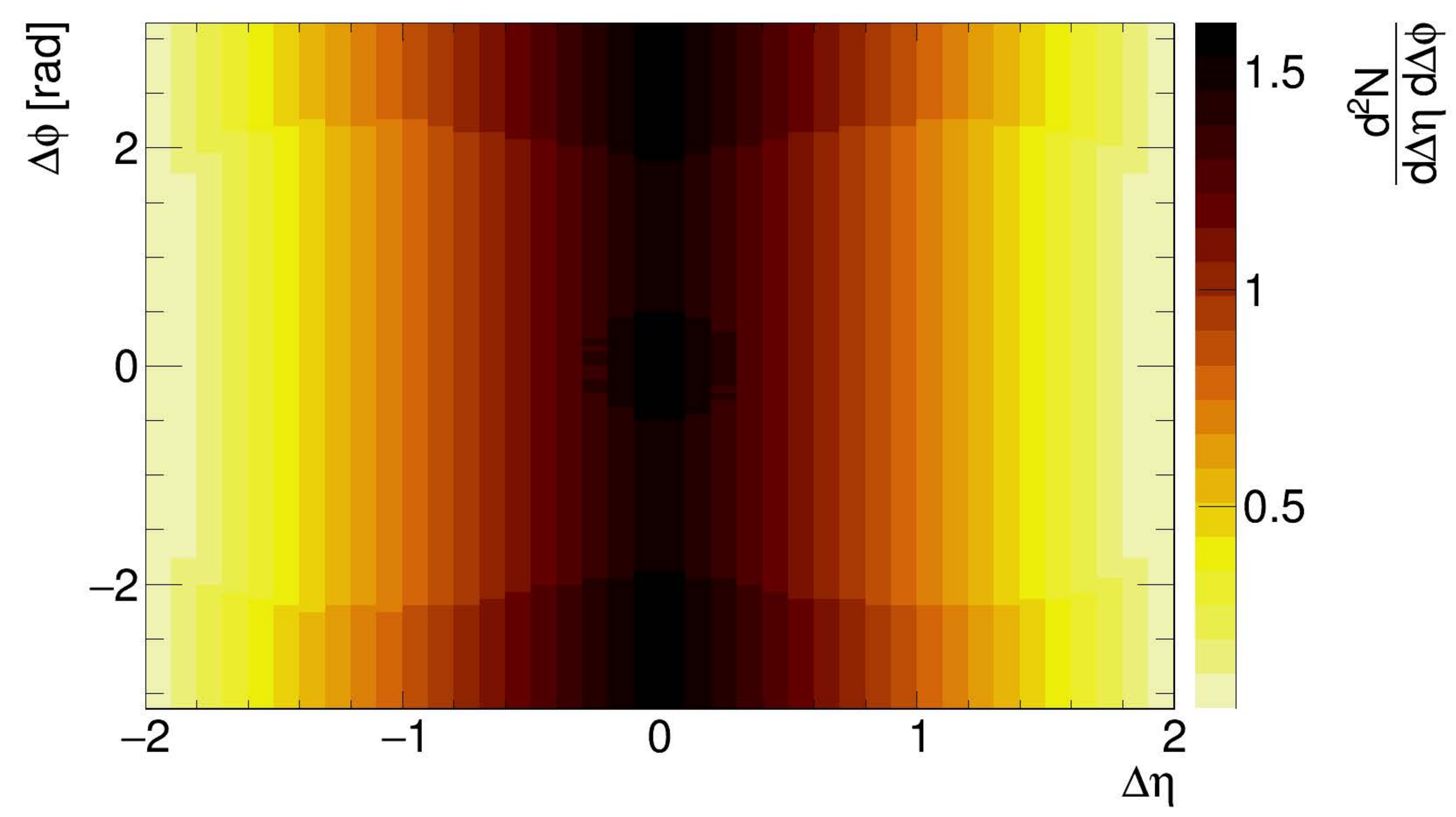}\hfill
	\caption{Two-particle ($\deta$, $\dphi$) distributions from diffractive (a) and non-diffractive (b) interactions. The two-particle ``triangle'' acceptance in $\deta$ is not corrected. The $z$ axis (color coordinate) is the number of pairs per unit of pseudorapidity per radian.}
	\label{fig:2Draw}
\end{figure}

Figure~\ref{fig:2Draw} shows $(\deta,\dphi)$ correlations of POI from diffractive and non-diffractive events. The two-particle acceptance, approximately a triangle in $\deta$ as discussed below, has not been corrected in these plots. The shapes are similar between diffractive and non-diffractive events, whereas the magnitude is significantly smaller in the former because of the smaller number of particles produced in diffractive events. 


Before continuing, we want to correct for an important aspect, namely, the two-particle acceptance. 
Consider a particle in a detector of limited acceptance in $\eta$, a second particle a distance $|\deta|$ away may fall outside the acceptance on one side. It would be helpful to correct for such a loss in acceptance. 
This acceptance is approximately a triangle in $\deta$ (for approximately uniform single-particle distribution in $\eta$), and is often referred to as the ``triangle" acceptance~\cite{Adams:2005ph}. 
We have assumed solenoidal detectors of full acceptance in $\phi$ (which is important for azimuthal anisotropy measurements), and because of the azimuthal periodicity, the two-particle $\dphi$ acceptance is   uniform.
We obtain the two-particle acceptance factor using the mixed-event technique, taking one particle from one event and the other from another event of the {\em same multiplicity value} as the first event. The mixed-event correlations are obtained separately for diffractive events and non-diffractive events. The mixed-event correlation function is normalized such that the magnitude is unity at $\deta=0$ because, with one particle within the $\eta$ acceptance (in our case $|\eta|<1$), the probability for the second particle to be inside the acceptance is 100\% if $\deta=0$. The normalized mixed-event correlation is divided to obtain the acceptance-corrected two-particle correlations.

\begin{figure}[H]
	\centering
    \xincludegraphics[width=0.48\textwidth,labelbox,label=a)]{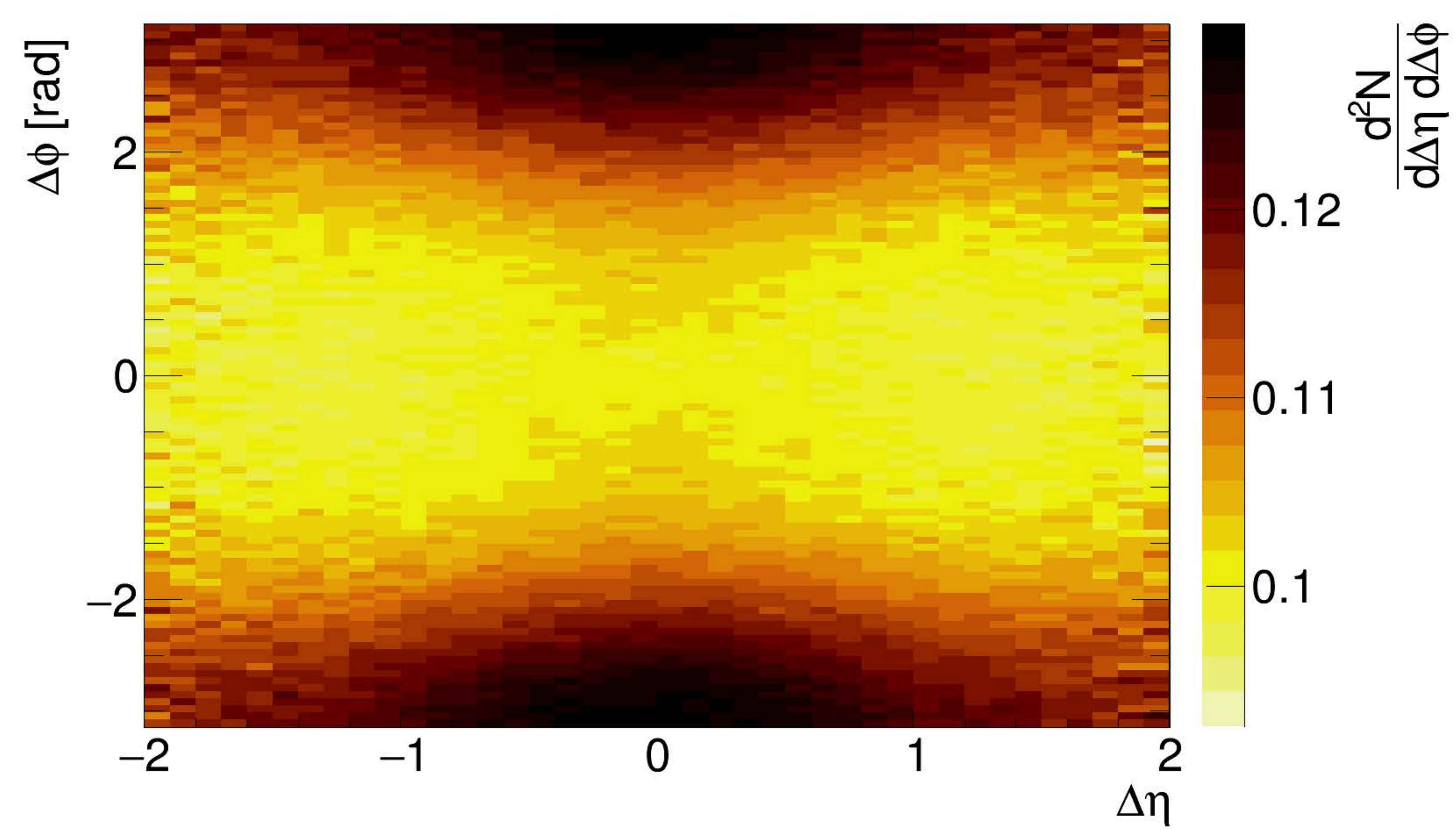}\hfill
	\xincludegraphics[width=0.48\textwidth,labelbox,label=b)]{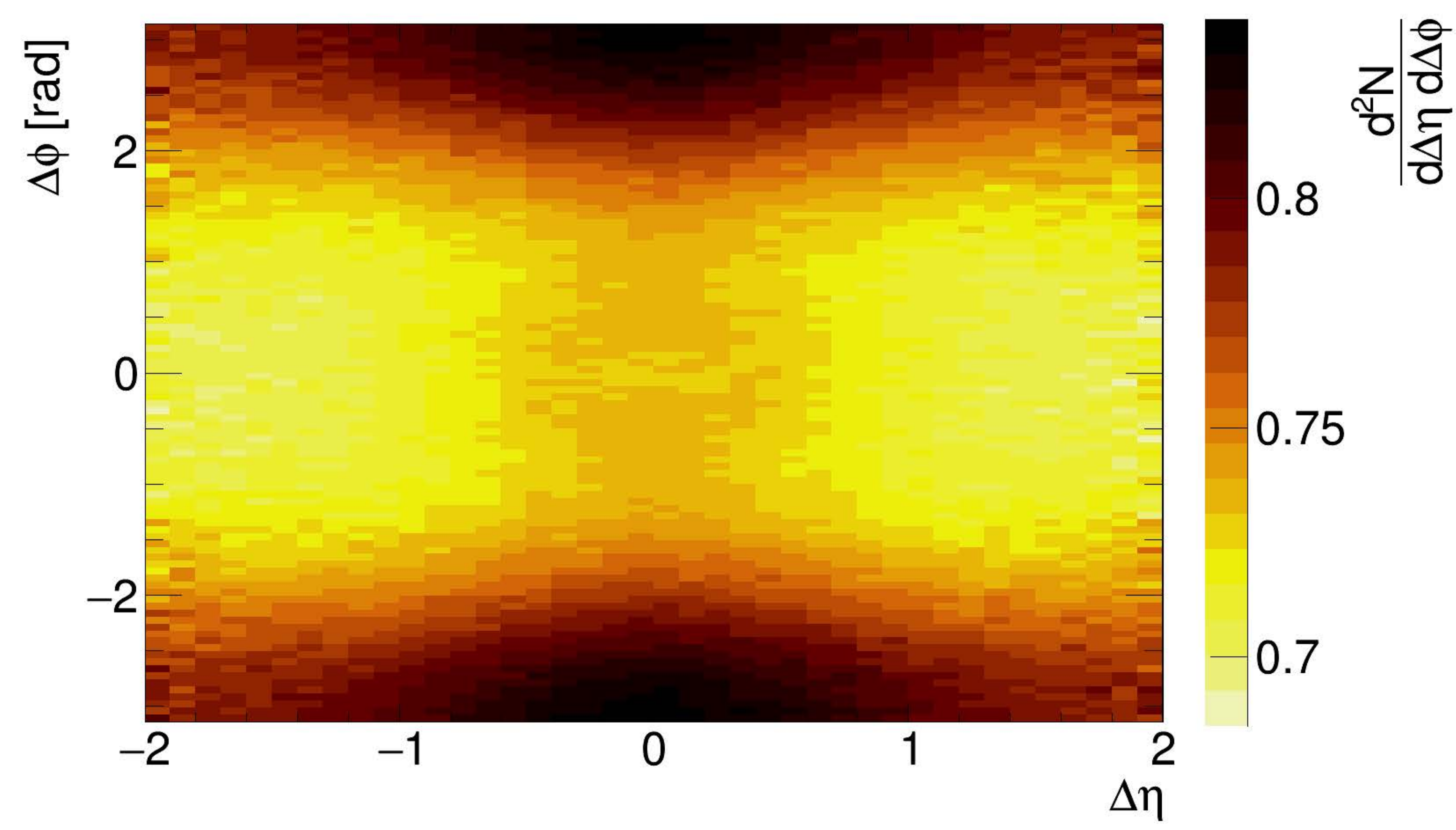}\hfill
    \xincludegraphics[width=0.48\textwidth,labelbox,label=c)]{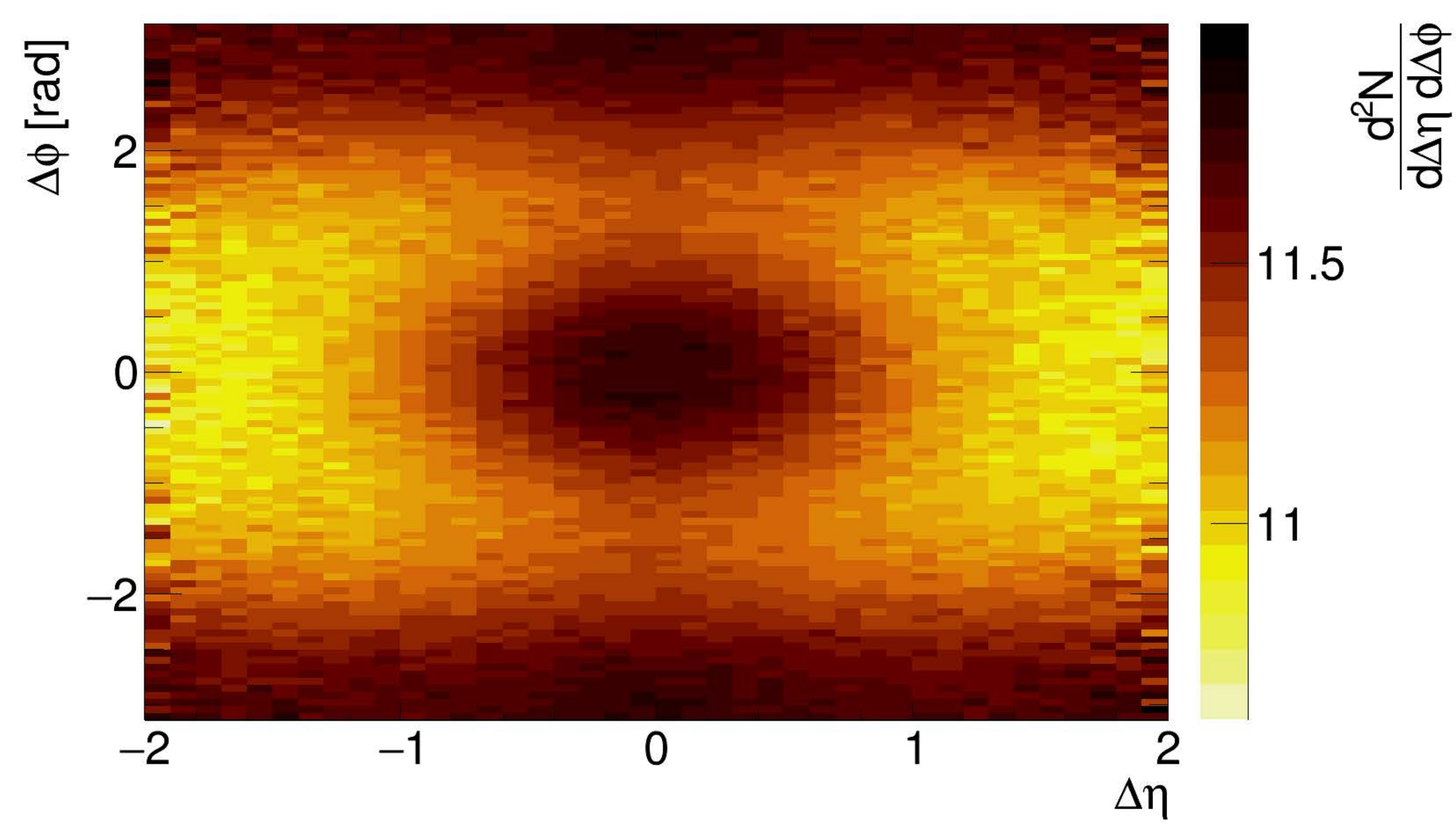}\hfill
	\caption{Two-particle correlations in non-diffractive interactions, as in Fig.~\ref{fig:2Draw}(b), in various multiplicity ranges of $1\leq\Nch\leq3$ (a), $4\leq\Nch\leq6$ (b), and $15\leq\Nch\leq20$ (c). 
    The two-particle ``triangle'' acceptance in $\deta$ is corrected by mixed events of the corresponding multiplicity range. The $z$ axis (color coordinate) is the number of pairs per unit of pseudorapidity per radian. For example, the $\deta$-acceptance corrected total number of pairs is approximately 20 in panel b), giving an average amplitude of approximately 0.8.}
	\label{fig:2D_125}
\end{figure}
Figure~\ref{fig:2D_125} shows the ``triangle" acceptance corrected $(\deta,\dphi)$ correlations in non-diffractive interactions for three multiplicity ranges. The correlations feature a near-side peak $(\deta\approx0,\dphi\approx0)$ (most prominent in high multiplicity events) and a broad away-side ridge at $\dphi\approx\pi$. These features are characteristic of intra-jet and inter-jet (dijet) correlations, as well as resonance decays and global momentum conservation. Correlations in diffractive interactions are similar. It is noteworthy that the near-side peak is not visible in low-multiplicity events, however, resonance decays (and similarly jet correlations) must be present in those events as well. As will be shown later, the negative $V_1$ component is large in those low-multiplicity events, such that the near-side correlation peak is submerged by the negative dip~\cite{Adamczyk:2015xjc}.

Far more resonances decay into daughters with an opposite-sign (OS) charge pair than a same-sign (SS) charge pair. Intra-jet correlated hadrons have more opposite-charge pairs than same-charge pairs because jets are mostly neutral gluon jets (i.e., charge-ordering phenomenon in jets). One thus has stronger correlation in OS pairs than in SS pairs, and can gain more insights by analyzing OS and SS correlations separately. 

\begin{figure}[H]
	\centering
    \xincludegraphics[width=0.48\textwidth,labelbox,label=a)]{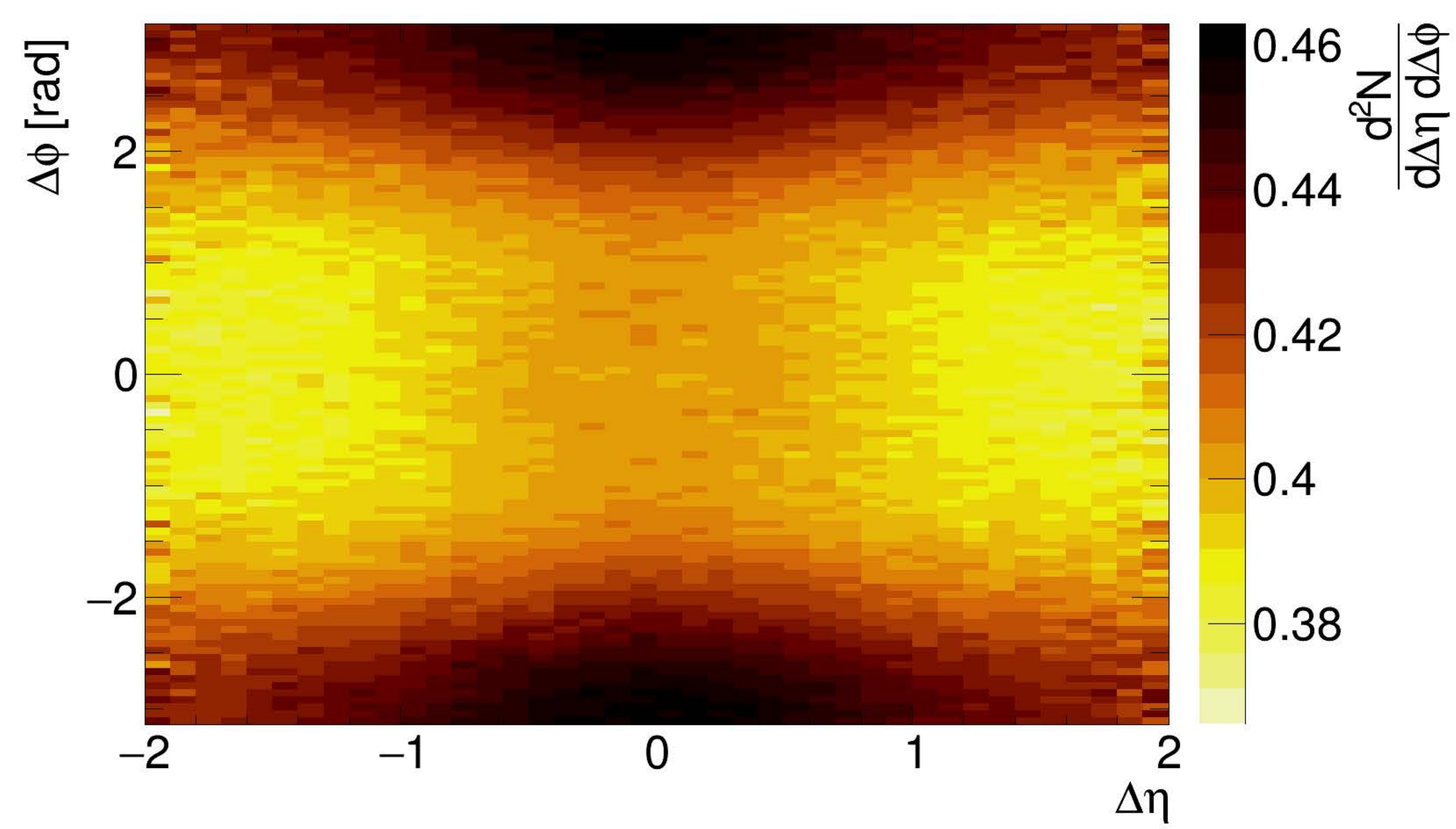}\hfill
	\xincludegraphics[width=0.48\textwidth,labelbox,label=b)]{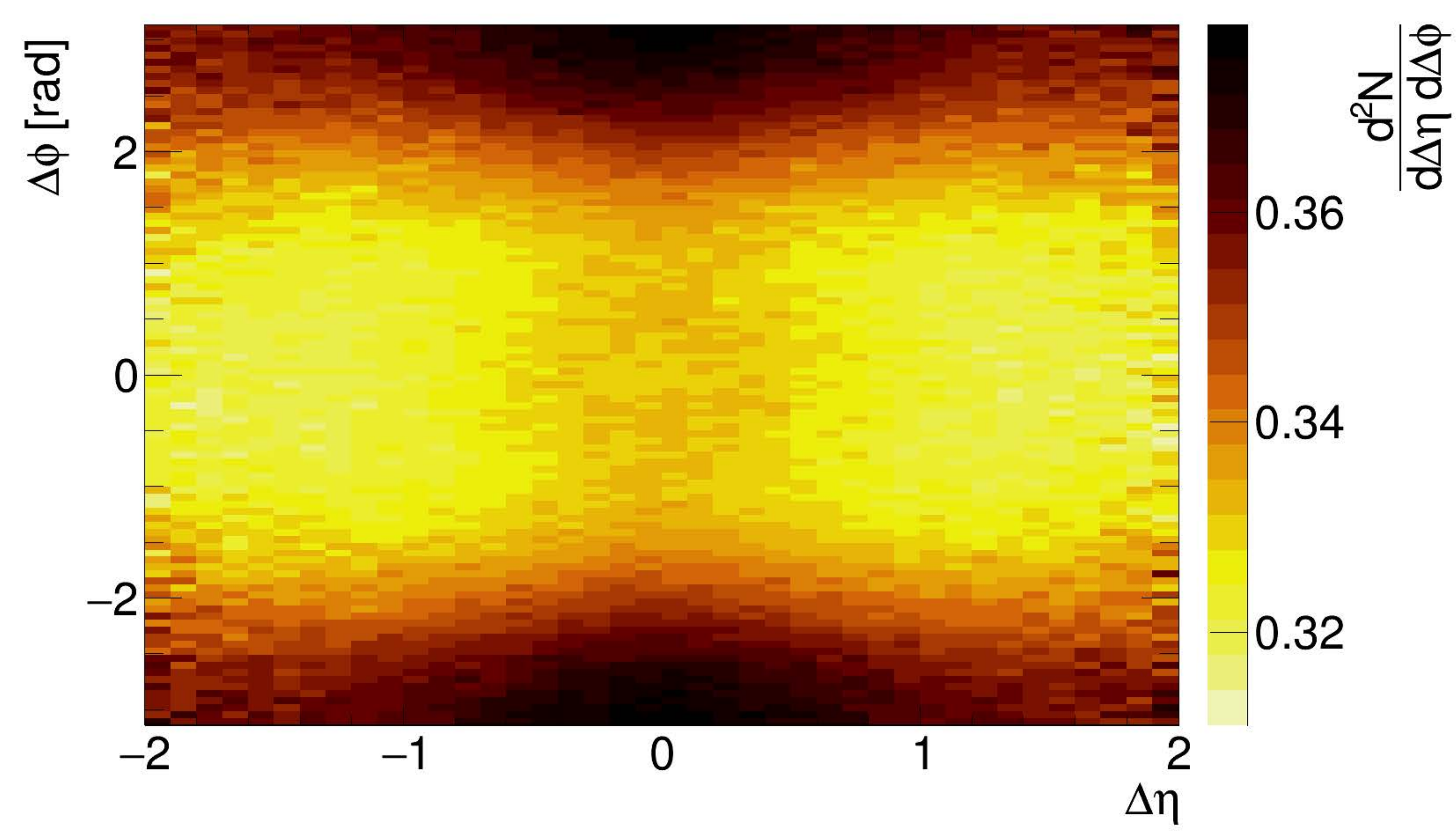}\hfill
	\caption{Acceptance-corrected two-particle $(\deta,\dphi)$ correlations in non-diffractive collisions with $4\leq\Nch\leq6$ as in Fig.~\ref{fig:2D_125}(b), but separately for OS pairs (a) and SS pairs (b). Correlations are similar in shapes for other multiplicity bins as well as for diffractive events.}
	\label{fig:2D}
\end{figure}
Figure~\ref{fig:2D} shows the acceptance-corrected correlations, separately for OS and SS pairs, in non-diffractive \pp\ collisions. One multiplicity bin is shown as an example; the results are similar in shape for other multiplicity bins. The near-side correlations are more prominent in OS correlations than in SS correlations, reflecting, presumably, resonance decays and charge ordering in (mini-)jet fragmentation. The away-side correlations ($\dphi\approx\pi$) are similar between OS and SS correlations, indicating no charge-dependent correlations on the away side.

\subsection{Charge-dependent correlations}
\begin{figure}[hbt]
    \xincludegraphics[width=0.24\textwidth,label=\hspace*{-0.2cm}a)]{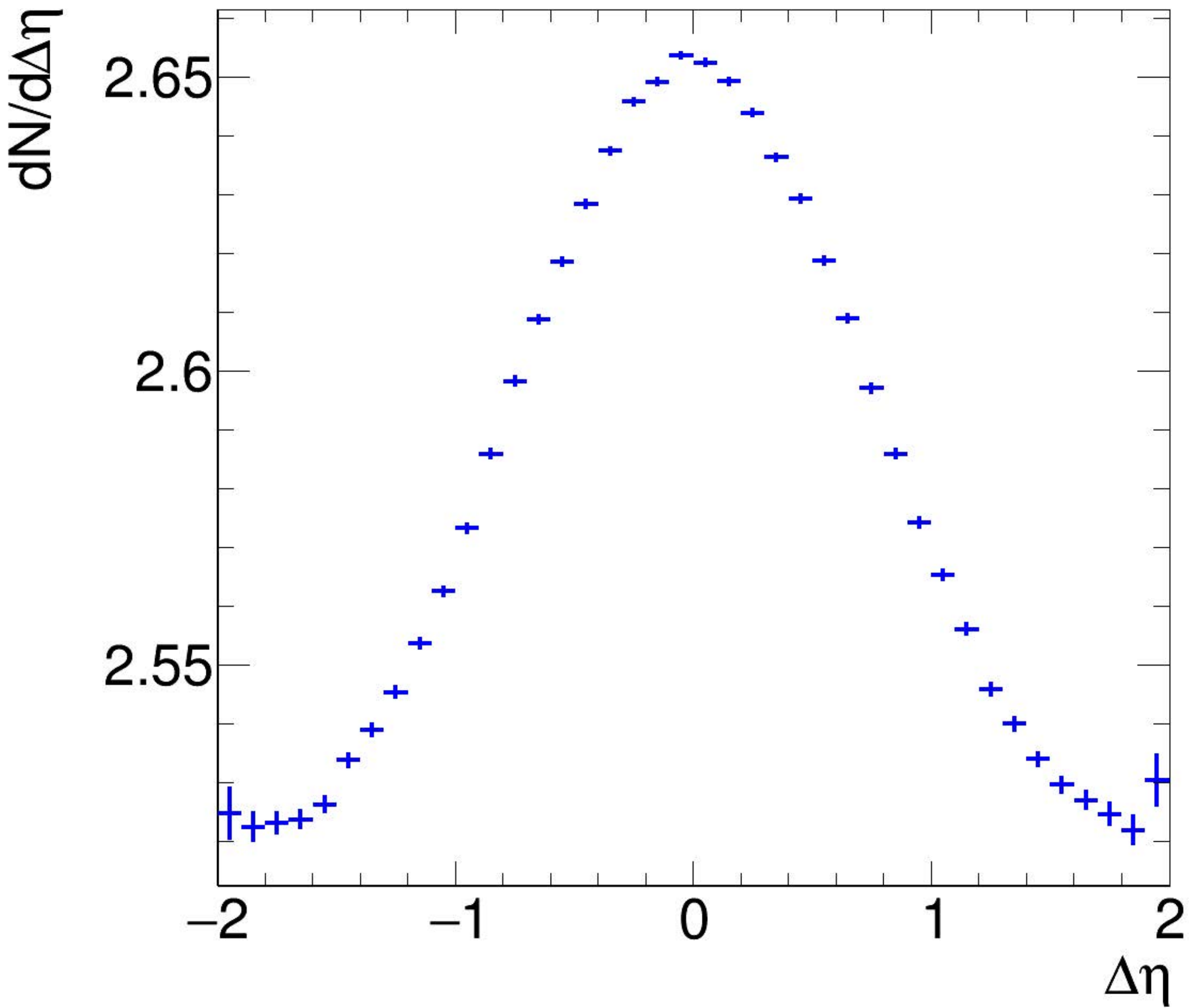}\hfill
	\xincludegraphics[width=0.24\textwidth,label=\hspace*{0.1cm}b)]{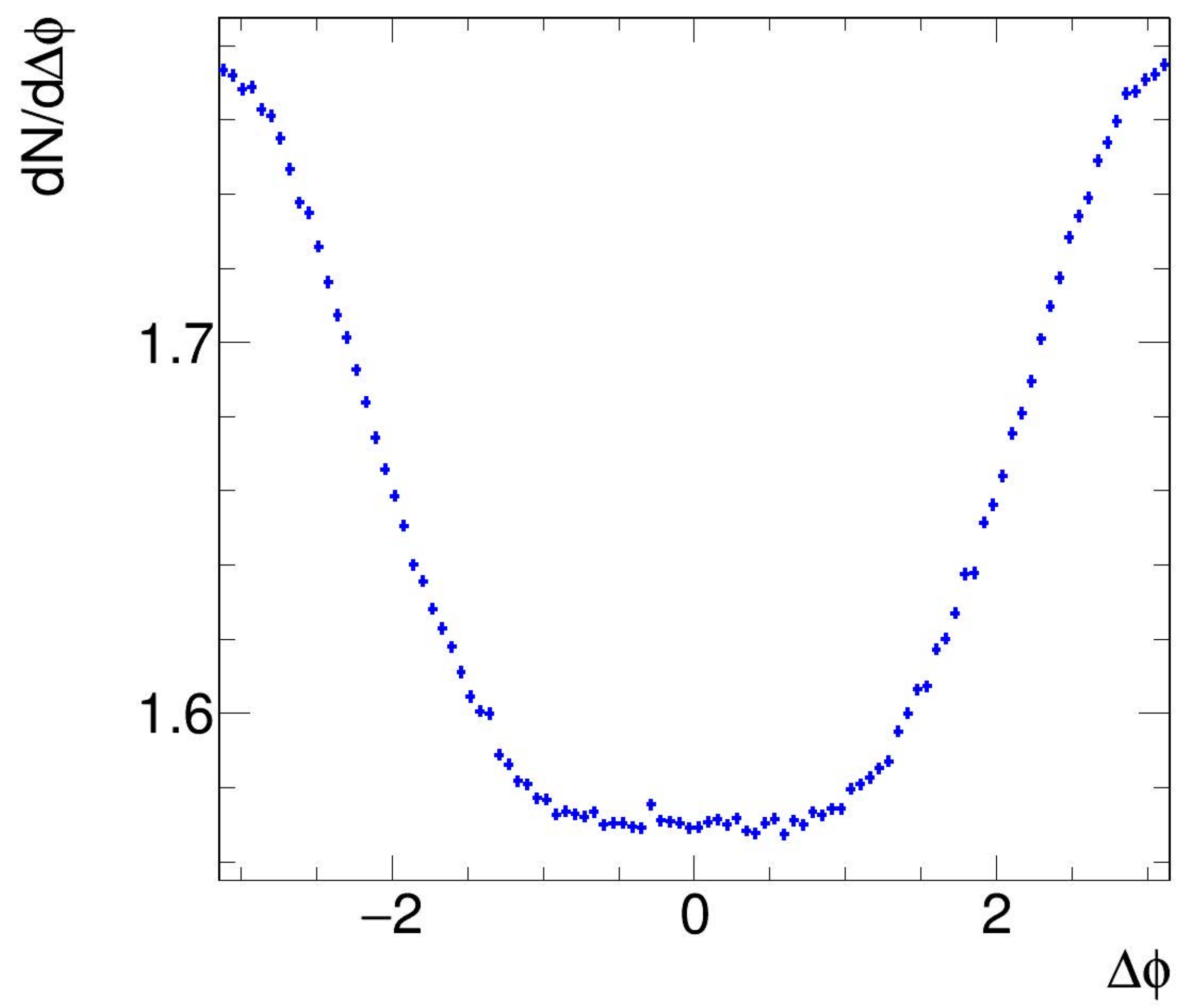}\hfill
	\xincludegraphics[width=0.24\textwidth,label=\hspace*{-0.2cm}c)]{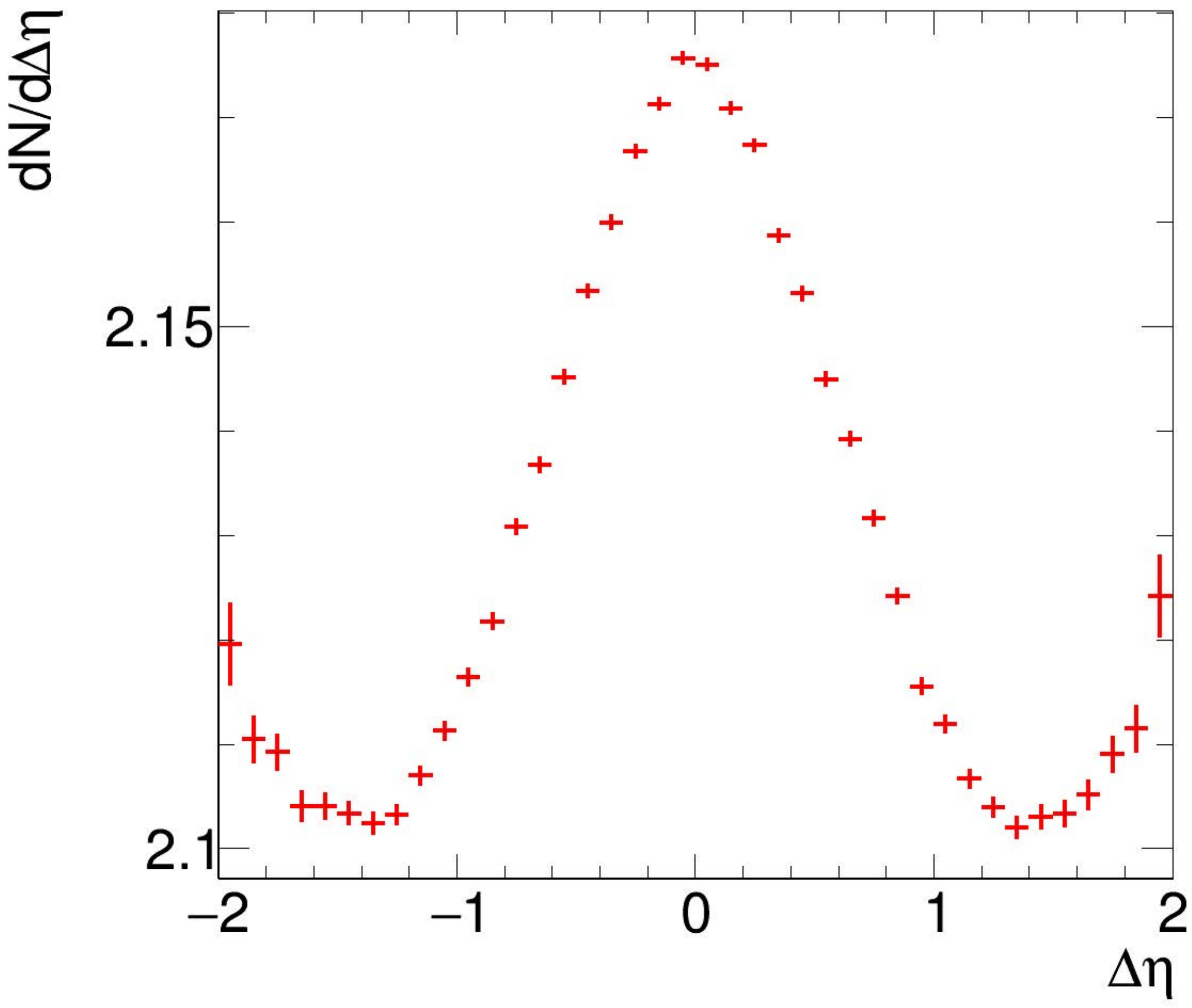}\hfill
	\xincludegraphics[width=0.24\textwidth,label=\hspace*{0.1cm}d)]{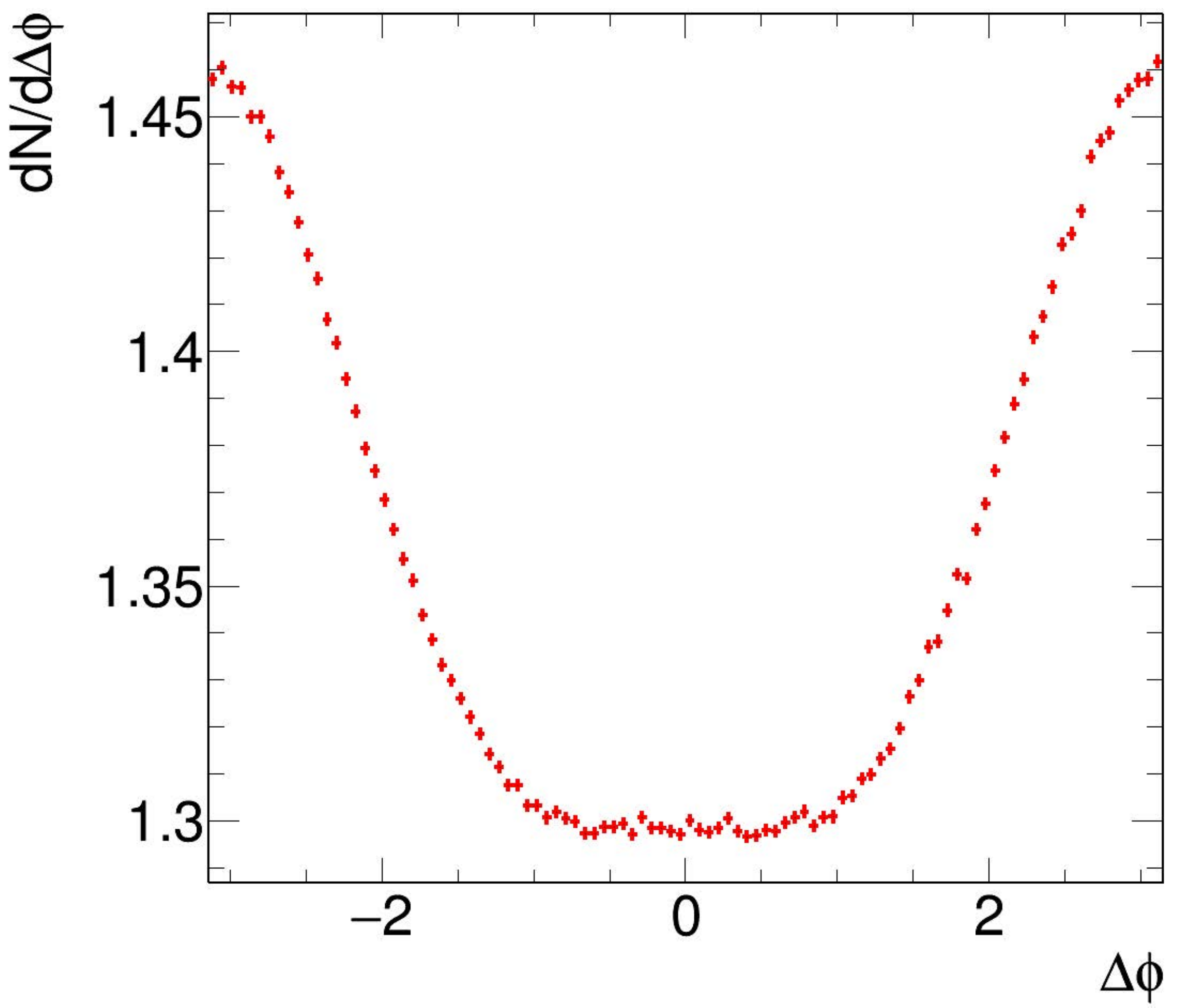}\hfill
	\caption{Projections of the 2D correlation functions from Fig.~\ref{fig:2D} onto $\deta$ (a,c) and $\dphi$ (b,d) for OS (a,b) and SS (c,d) pairs in non-diffractive events with $4\leq\Nch\leq6$. The correlations are normalized per event and by bin size.}
	\label{fig:proj}
\end{figure}

Figure~\ref{fig:proj} shows the $\deta$ and $\dphi$ projections of the OS and SS 2D correlation functions for non-diffractive events with $4\leq\Nch\leq6$ from Fig.~\ref{fig:2D}. Other multiplicity bins are similar in the correlation shapes. The $\deta$ correlation is characteristic of a Gaussian centered at $\deta=0$. The $\dphi$ correlations are dominated by the away-side peak at $\dphi=\pm\pi$, with essentially no near-side peak at $\dphi=0$ visible at the scale of the plots because of the aforementioned large negative dipole. 

It is known that the OS correlation contains more contributions from resonance decays and jets than the SS correlation, because of contributions from neutral resonances and charge ordering in jet fragmentation. The number of OS and SS pairs from the underlying event also differ significantly as shown in Fig.~\ref{fig:proj}. It requires careful analysis to study those charge-dependent nonflow correlations. We leave it to future work. In this study, we are interested in the possible flow-like anisotropies in \pp\ events simulated using  \pythia. Because the SS pairs are less contaminated by nonflow, we  focus on SS correlations in the following sections.

\subsection{Same-sign correlations}
Although to a less degree than the OS pairs, SS pairs are affected by jet and dijet correlations as well. Intra-jet correlations are concentrated on the near side $\dphi\sim0$. The dijet correlations are back-to-back and concentrated on the away side $\dphi\sim\pm\pi$. Because the underlying parton-parton interaction kinematics are not necessarily symmetric about midrapidity, the away-side correlations are approximately $\deta$ independent. Other back-to-back correlations, such as global momentum conservation and possibly flow-like harmonics, are also mostly independent of $\deta$. One may thus take the difference between small- and large-$\deta$ correlations to assess the near-side nonflow correlations.

\begin{figure}[hbt]
	\centering
    \xincludegraphics[width=0.4\textwidth,label=\hspace*{0.9cm}a),pos=nwlow]{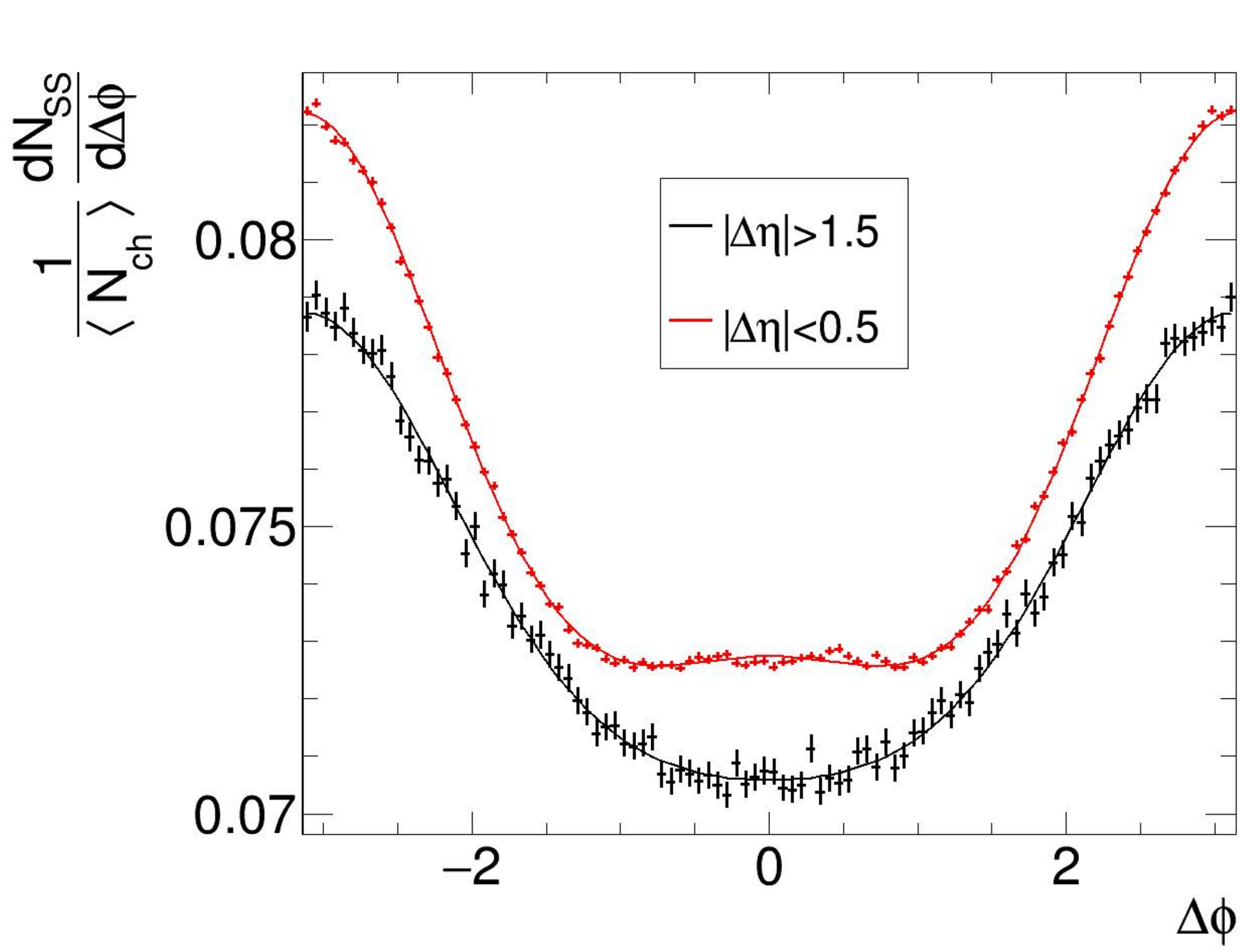}
    \xincludegraphics[width=0.408\textwidth,label=\hspace*{0.9cm}b),pos=nwLow]{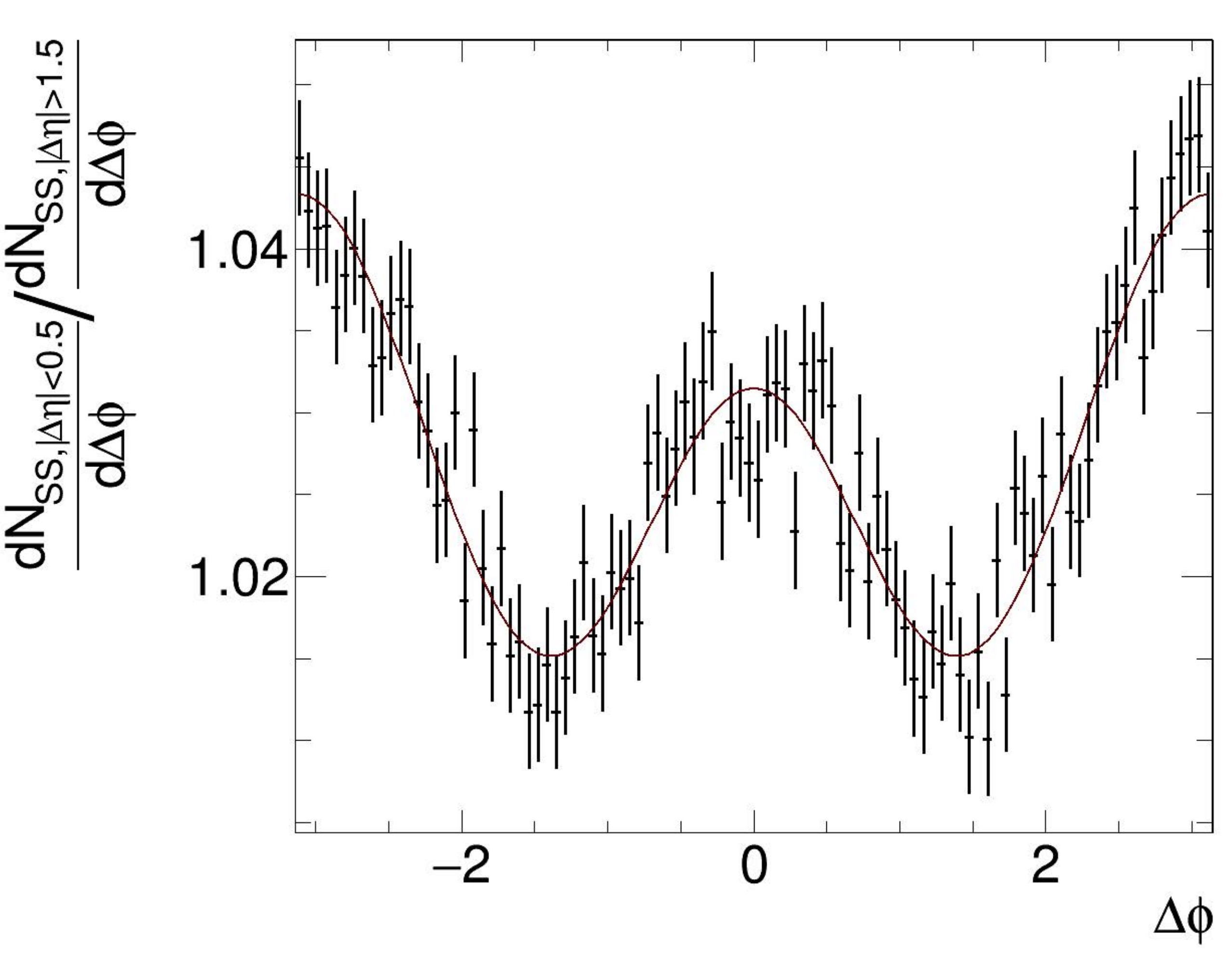}
	\caption{(a) SS correlations in $\dphi$ within small $|\deta|<0.5$ and large $1.5<|\Delta\eta|<2$ for non-diffractive events with $4\leq\Nch\leq6$. The correlations are obtained from projections of the acceptance-corrected $(\deta,\dphi)$ correlations in Fig.~\ref{fig:2D} in corresponding $\deta$ ranges, and are then divided by $\mean{\Nch}$; therefore, they are SS dihadron correlations per hadron.  Each superimposed curve is a fit to the corresponding data with the sum of a pedestal and two Gaussians, one centered at $\dphi=0$ and the other at $\dphi=\pm\pi$. The fitted pedestals are 
    $0.0708\pm 0.0007$ and $0.0704\pm 0.0001$ 
    for small and large $\deta$, respectively. The central Gaussian amplitude for the large-$\deta$ correlation is negative. 
    (b) Ratio of small-$\deta$ correlation over large-$\deta$ one. The superimposed curve is a fit to the same functional form as in panel a). The fitted pedestal value is $0.96\pm 0.06$. 
    }
    \label{fig:small_large_deta}
\end{figure}
We define small $\Delta \eta$ to be $|\Delta \eta| < 0.5$ and large $\Delta \eta$ to be $1.5 < |\Delta \eta| < 2$ (the upper limit is automatically satisfied because of our single-particle acceptance requirement of $|\eta|<1$). Note that we have defined an equal $|\deta|$ size for the two regions, but the specific $\deta$ values for the definitions are less important as long as the large-$\deta$ region contains minimal near-side contribution.
Figure~\ref{fig:small_large_deta}(a) shows the $\dphi$ correlations at small and large $\deta$ for the medium multiplicity range, as an example.
The $\Delta\eta$ acceptance has already been corrected in those correlations. 
The small- and  large-$\deta$ correlations are similar in amplitude, suggesting a more or less uniform two-particle distributions in $\deta$.
The correlations are fit with the functional form of two Gaussians atop a constant pedestal, with one Gaussian centered at $\dphi=0$ and the other at $\dphi=\pm\pi$. The fitted pedestal values are approximately equal between them.
Figure~\ref{fig:small_large_deta}(b) shows the ratio of the correlations between small and large $\deta$ values. The ratio is fit with the same functional form as above, and the fitted pedestal value is $0.96\pm 0.06$.

\begin{figure}[h!]
	\centering
	\includegraphics[width=0.45\textwidth]{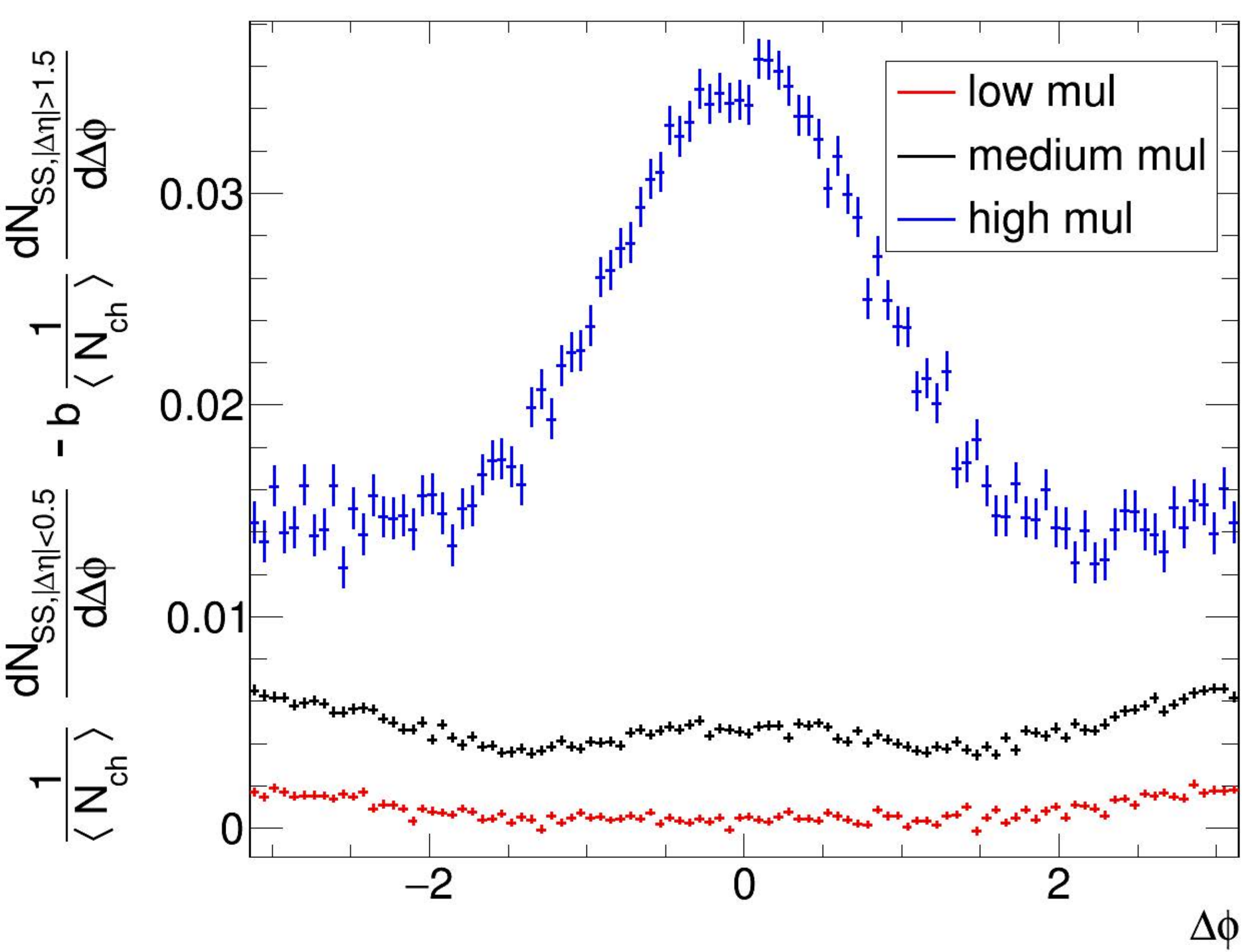}
 	\caption{The differences between small- and large-$\Delta\eta$ dihadron correlations for three multiplicity ranges of $1\leq\Nch\leq3$, $4\leq\Nch\leq6$, and $15\leq\Nch\leq20$, where the large-$\Delta\eta$ correlation is first scaled by a normalization factor. 
    Those correlations for the medium multiplicity range are shown in Fig.~\ref{fig:small_large_deta}(a).
    The normalization factor is taken to be the pedestal value of the corresponding ``pedestal + two-Gaussian'' fit, such as the one in the lower panel of Fig~\ref{fig:small_large_deta}(b); the pedestal values for $1\leq\Nch\leq3$ and $15\leq\Nch\leq20$ (that are not shown in Fig.~\ref{fig:small_large_deta}) are $0.95 \pm 0.01$ and $0.96 \pm 0.07$, respectively.} 
    \label{fig:nearside}
\end{figure}

Figure~\ref{fig:nearside} shows the difference between small- and large-$\Delta\eta$ dihadron correlations, with the large-$\deta$ correlation first multiplied by the fitted pedestal to the ratio in Fig.~\ref{fig:small_large_deta}(b). 
Because of this choice of normalization, the correlation difference is not necessarily zero outside the near-side region. If we normalized the small- and large-$\Delta\eta$ correlations by matching their away side,  the correlation differences would all be averaged at zero outside the near-side region.
For the low-multiplicity bin, there is no near-side peak because of the large negative dipole and likely minimal intra-jet correlations. The near-side peak begins to show up in the mid-multiplicity bin, and is predominant for the high-multiplicity bin, suggesting strong contributions from intra-jet correlations. Resonance decay contributions are likely minimal in SS pairs.

\subsection{Large $\Delta \eta$ dihadron azimuthal correlations\label{sec:large_deta}} 

\begin{figure}
	\centering
	\includegraphics[width=0.46\textwidth]{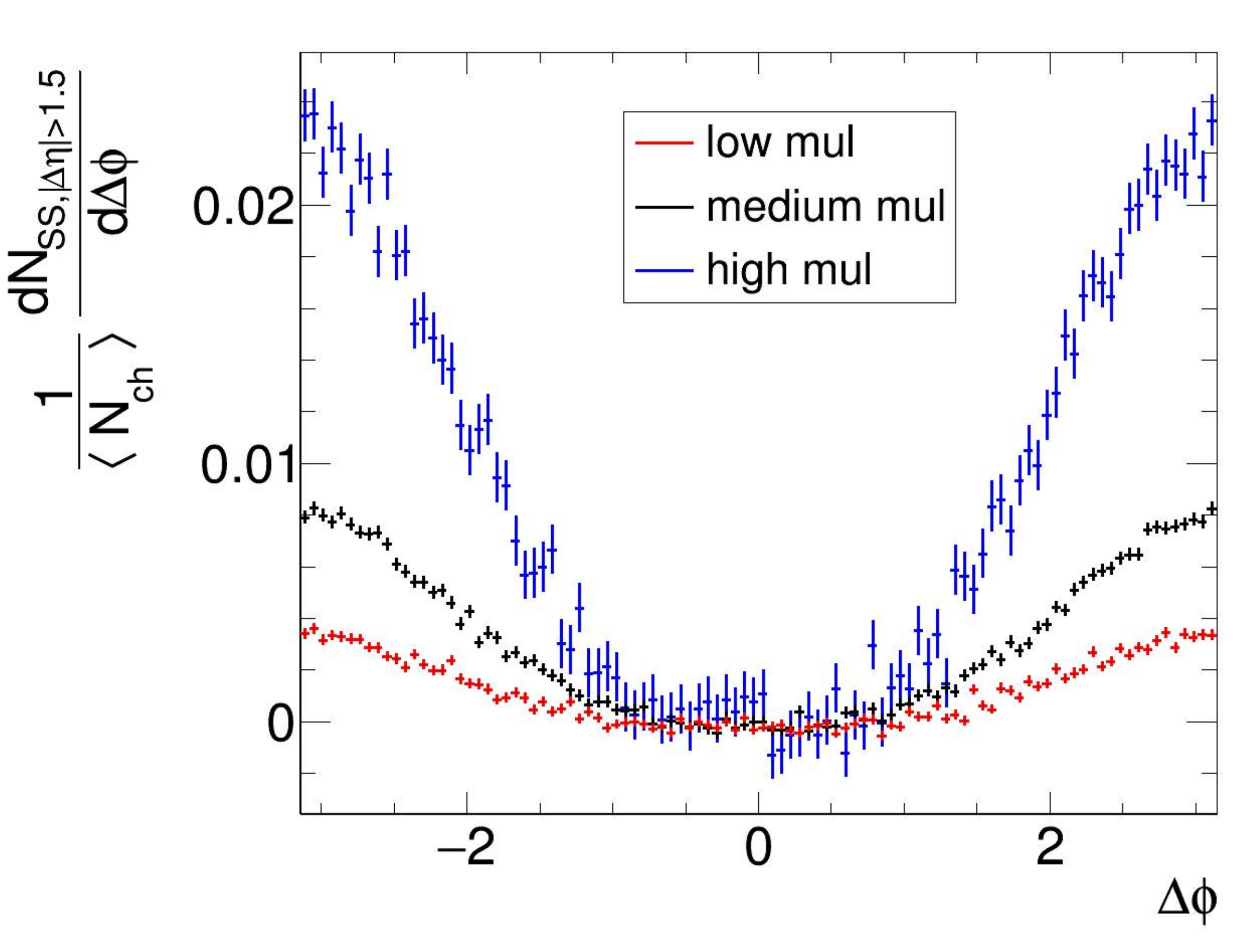}
	\caption{The SS per-trigger dihadron $\Delta \phi$ correlations at large $\deta$ ($1.5<|\deta|<2$) in non-diffractive \pp\ collisions simulated using  \pmodel, for three multiplicity ranges $1 \leq \Nch \leq 3$ (low mult.), $4\leq\Nch\leq6$ (medium mult.), and $ 15 \leq \Nch \leq 20$ (high mult.). Each distribution is subtracted by the constant pedestal from the ``pedestal + Gaussian'' fit, like in Fig.~\ref{fig:small_large_deta} for the medium-multiplicity range. The subtracted pedestals are $0.035 \pm 0.010$, $0.07 \pm 0.0001$, and $0.342 \pm 0.001$ for the three multiplicity ranges, respectively. } 
    \label{Proj_no_ped}
\end{figure}

Intra-jet correlations are insignificant at large $\deta$;
therefore, it is cleaner to extract flow anisotropies from
large-$\deta$ correlations, although dijet correlations are still present on the away side.
Figure~\ref{Proj_no_ped} depicts the large-$\deta$ per-trigger dihadron correlations for the three selected multiplicity bins. 
For easy display, the correlations are chopped off a constant pedestal such that the near-side region is zero for all three multiplicity ranges. The pedestal is obtained from a ``pedestal + Gaussian'' fit where the Gaussian describes the away-side correlation about $\dphi=\pm\pi$.

The correlation shapes are similar, with a prominent away-side peak and minimal near-side correlations. The away-side peak is primarily of a negative dipole shape ($-\cos\dphi$); this is in line with the expectation from global momentum conservation~\cite{Borghini:2006yk} and with the expectation that back-to-back dijet correlations at low $\pt$ values are also primarily of a negative dipole shape.
The away-side peak amplitude increases with the event multiplicity. This appears to suggest that the back-to-back dijet contribution increases with the event multiplicity, as global momentum conservation should not depend on multiplicity. This is likely owing to multiplicity bias--when high-multiplicity is demanded, we likely select events with relatively stronger jet contributions~\cite{PHENIX:2013jxf,Adamczyk:2014fcx,ALICE:2018ekf}.

\subsection{Model fits}
With the understanding that the SS correlations are composed of intra-jet and dijet correlations atop  an underlying event with possible harmonic oscillations, we proceed to fit the mixed-event acceptance corrected $(\deta,\dphi)$ correlations. 
We use the following fit function:
\begin{widetext}
\begin{equation}\label{eq:fit}
    \frac{d^2N}{d\deta d\dphi} = 
    Ae^{- \frac{(\deta)^2}{2\sigma_{\eta}^2}} \left( e^{- \frac{(\dphi)^2}{2\sigma_{\phi}^2}} + e^{- \frac{(\dphi + 2\pi)^2}{2\sigma_{\phi}^2}} + e^{- \frac{(\dphi - 2\pi)^2}{2\sigma_{\phi}^2}} \right) +
    A_{\rm RG} e^{- \frac{(\deta)^2}{2\sigma_{\rm RG}^2}} + 
    \frac{B}{2 - |\Delta \eta|} {\rm erf} \left( \frac{2 - |\Delta \eta|}{b} \right) + 
    C \left(1 + 2\sum_{n=1}^{4}V_n\cos(n\dphi) \right)\,.
\end{equation}
\end{widetext}
The first term on the r.h.s.~of Eq.~(\ref{eq:fit}) describes the near-side 2D Gaussian in $(\deta,\dphi)$; the repeated $\dphi$-Gaussian at $\pm2\pi$ ensures azimuthal periodicity with those beyond $2\pi$ safely neglected. 
The second term describes the $\dphi$ ridge that is present as a Gaussian at $\deta=0$; such a ridge could be the result of the aforementioned beam jet fragmentation~\cite{Andersson:1983ia}.
The third term attempts to describe a feature present in the correlation functions that bends up towards large $|\deta|$, which can be attributed to back-to-back emission of particles in $\eta$ due to, e.g., {\em local} momentum conservation; enabling $\eta_{1}+\eta_{2}$ to fluctuate about zero in Gaussian and integrating it within the acceptance boundaries of $|\eta|<1$ results in the marginal distribution of an error function in $\Delta\eta$.
The fourth term is the underlying event contribution with possible harmonic oscillations; we keep only the first four harmonics in the fit, where the $V_n$ $(n=1,2,3,4)$ quantify their amplitudes. Note that these harmonic anisotropy amplitudes are with respect to the underlying event multiplicity quantified by the baseline parameter $C$, excluding contributions from the nonflow (intra-jet and dijet) particles. 
It is found that no additional structure is required on the away side to describe the away-side jet. This is because the away-side jet contribution, in our $\pt$-inclusive study (dominated by low $\pt$ particles), can presumably be described by a negative $V_1$ dipole and blended in the $V_1$ harmonic term.

\begin{figure}[H]
	\centering

 \includegraphics[width=0.48\textwidth]{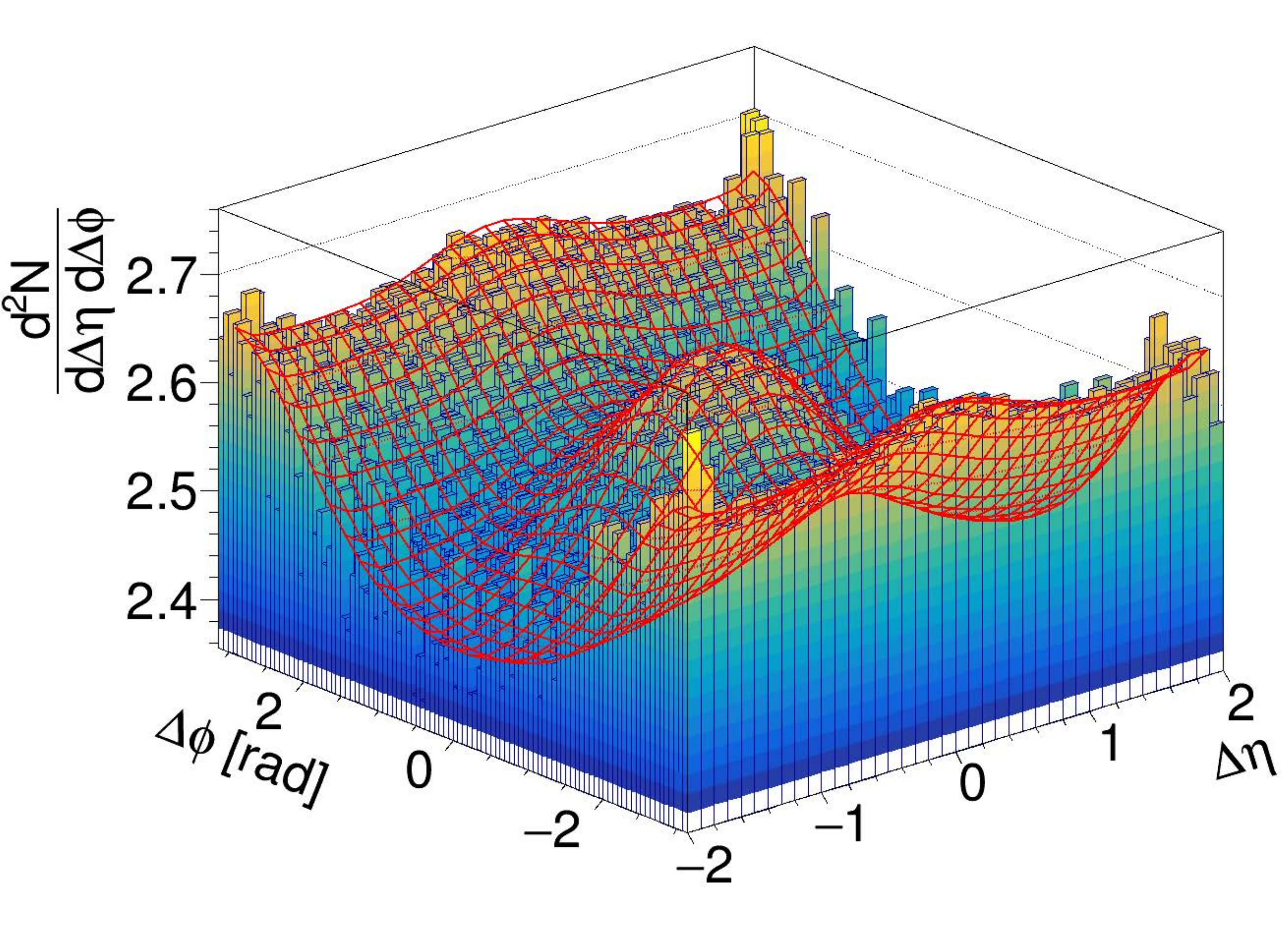}\hfill
	\caption{An example of the 2D model fit by Eq.~(\ref{eq:fit}) to the SS pair correlations for the multiplicity range $11\leq\Nch\leq14$.}
	\label{fig:fit}
\end{figure}

We fit the $(\deta,\dphi)$ SS correlation in each multiplicity bin using Eq.~(\ref{eq:fit}). Figure~\ref{fig:fit} shows, as an example, the fit result for the $7\leq\Nch\leq10$ multiplicity bin of the non-diffractive events. The $\chi^2/{\rm NDF}$ values are 1.24, 1.54, 1.12, 1.06, and 0.99 for the multiplicity bins of 1--3, 4--6, 7--10, 11--14, and 15--20 of the non-diffractive events, respectively. For diffractive events, the $\chi^2/{\rm NDF}$ values are 1.78, 1.80, 1.08, and 1.01 for the multiplicity bins of 1--3, 4--6, 7--10, and 11-14, respectively. The number of degrees of freedom (NDF) is 3988 for all correlation functions. 
We note that the $\chi^2/{\rm NDF}$ values are not optimal, suggesting that the fit function of Eq.~(\ref{eq:fit}) is not ideal. Nevertheless, the fitted $V_n$ parameters may still provide a reasonable description of the underlying modulation. The fit quality can be improved by introducing more components (parameters) into the fit function; however, such improvements may be at the expense of clarity. 

\section{Anisotropy Results and Discussions}

\begin{figure}[H]
	\centering
	\includegraphics[width=0.45\textwidth]{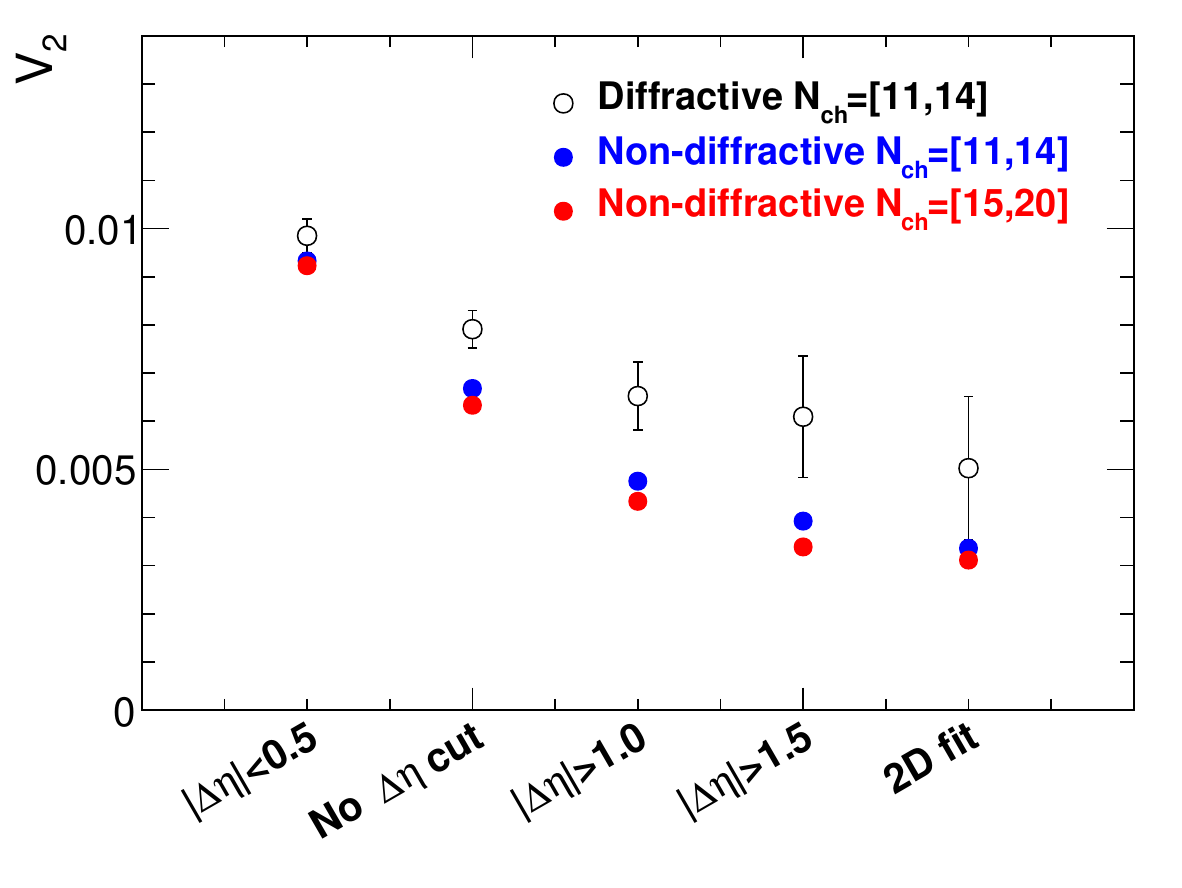}
	\caption{The $V_2$ values obtained from SS correlation functions by various methods. Shown are the results for diffractive events of one multiplicity bin and non-diffractive events of two multiplicity bins. The rightmost set of data points are from 2D fits. The rest are Fourier coefficients of correlation functions with various $\deta$ cuts. }
	\label{fig:V2_vs_method}
\end{figure}

We are most interested in the $V_n$ parameters. Multiple methods can be used to obtain $V_n$, whose physical interpretation depends on the method used~\cite{Feng:2024eos}. It can be obtained from the two-particle cumulant by the Fourier coefficient of Eq.~(\ref{eq:Fourier}) as well as by the fitting method of Eq.~(\ref{eq:fit}). Figure~\ref{fig:V2_vs_method} shows the $V_2$ obtained by various methods from the SS pair correlations for a selected multiplicity bin of diffractive events and two multiplicity bins of non-diffractive events: the leftmost data point is the Fourier coefficient of SS correlations within $|\deta|<0.5$, which gives a large value because of nonflow contribution from the near-side peak; the second data point from the left is that without any $\deta$ cut; the third point is that with $|\deta|>1.0$ and the fourth one is that with $|\deta|>1.5$, both of which yield a significantly reduced value because of the exclusion of the main part of nonflow at small $|\deta|$; the rightmost data point is the $V_2$ obtained from the 2D fit, which is the smallest for the results shown, suggesting a large removal of nonflow effects. 

\begin{figure}[hbtp]
	\centering
    \xincludegraphics[width=0.4\textwidth,pos=nw,label=\hspace*{3.7cm}a)]{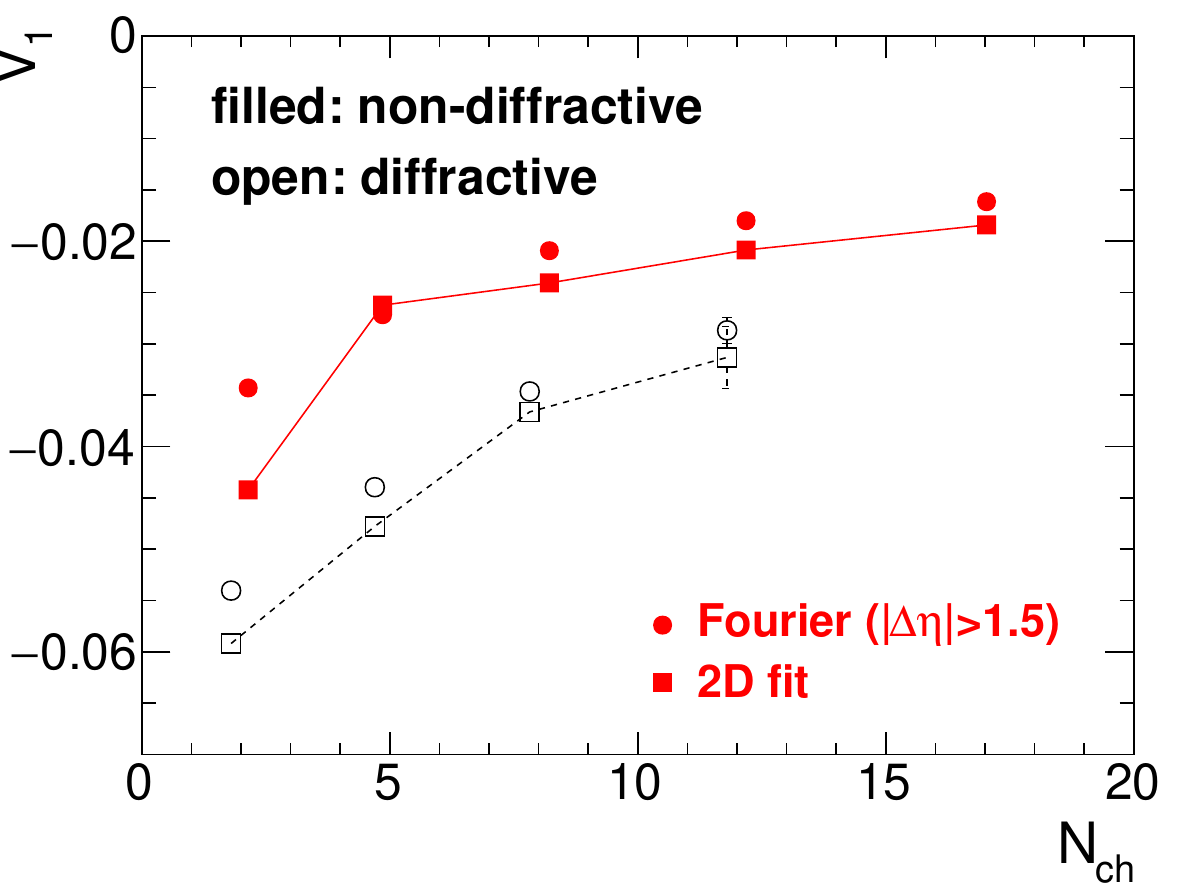}\hfill
    \xincludegraphics[width=0.4\textwidth,pos=nw,label=\hspace*{3.7cm}b)]{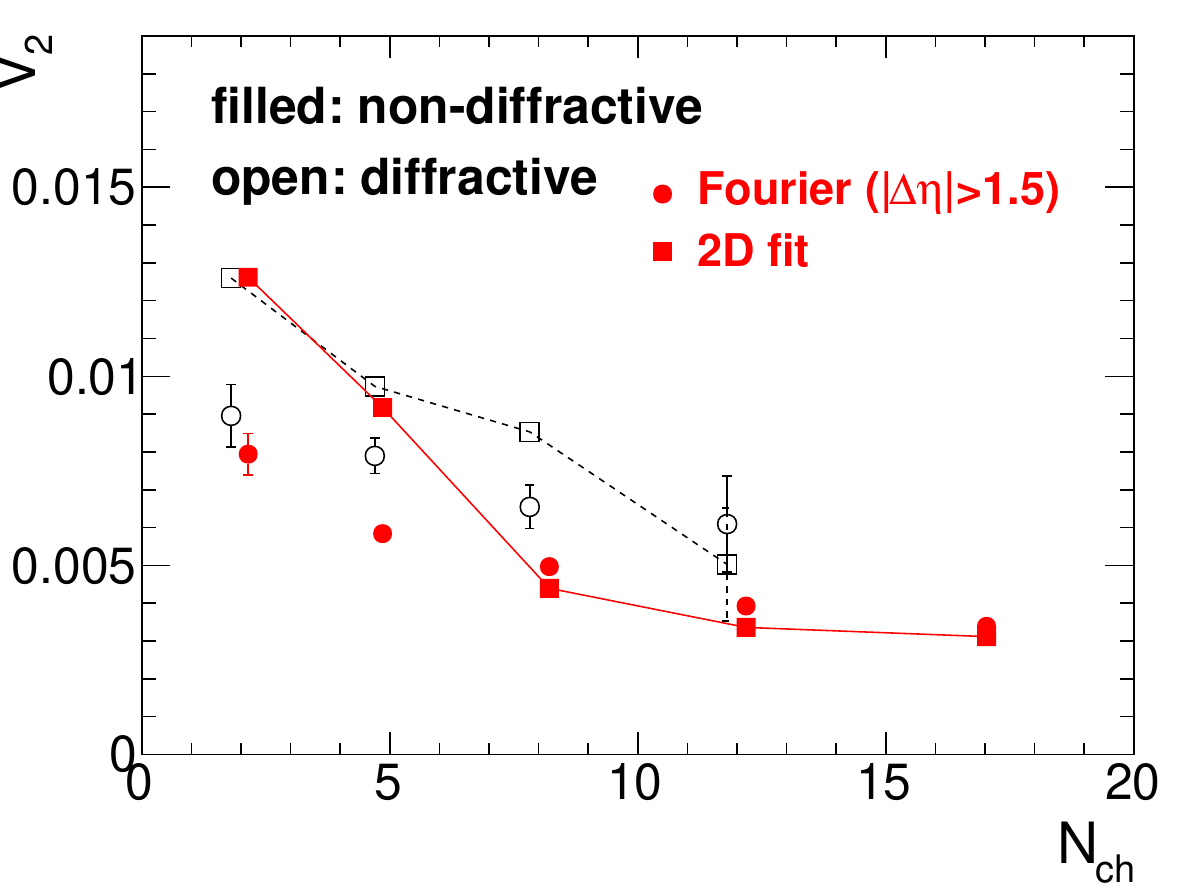}\hfill
    \xincludegraphics[width=0.4\textwidth,pos=nw,label=\hspace*{3.7cm}c)]{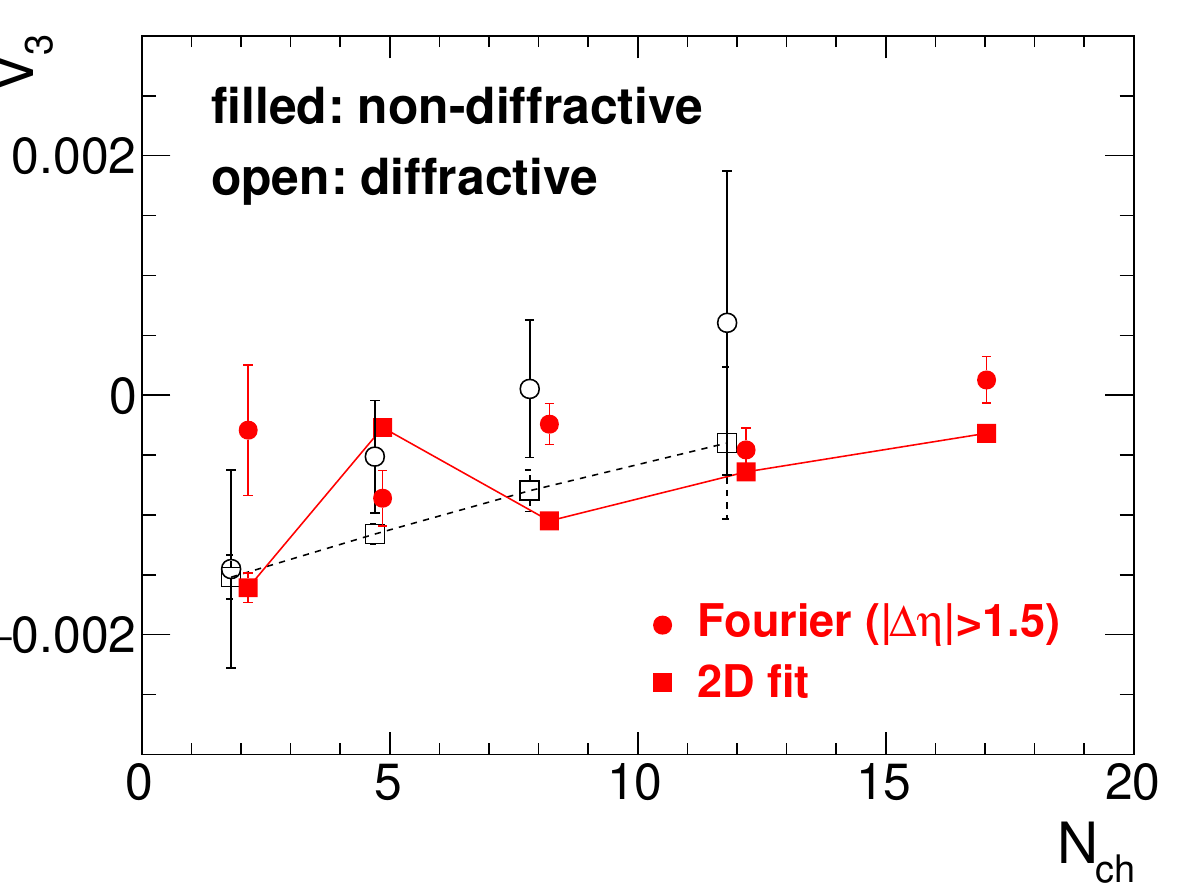}\hfill
	\caption{The $V_n$ parameters obtained from the 2D fit (circles, connected by lines), together with the Fourier coefficients calculated from the SS correlations at $1.5<|\deta|<2$ (squares). Results are shown for diffractive (open black points) and non-diffractive (filled red points) events as functions of multiplicity.}
	\label{fig:Vn}
\end{figure}

Figure~\ref{fig:Vn} shows $V_1$, $V_2$, and $V_3$ values obtained from the Fourier coefficients of the SS correlations at $|\deta|>1.5$ (such as the one in Fig.~\ref{fig:small_large_deta}(a)) and from the 2D fit to the SS $(\deta,\dphi)$ correlations. These two methods should have the smallest nonflow contamination. 
The values are plotted as functions of $\Nch$ for both non-diffractive (open black points) and diffractive events (filled red points). 
We can infer from Fig.~\ref{fig:Vn}(a) that the negative dipole $V_1$ amplitude decreases with increasing multiplicity. However, the multiplicity dependence is not $1/\Nch$, which would be characteristic of global momentum conservation~\cite{Borghini:2006yk}. Thus, the results suggest that the $V_1$ component in those \pp\ interactions is not of the nature of global momentum conservation. 
It is interesting to note that the $V_1$ amplitude is larger in diffractive events than in non-diffractive events at similar multiplicities.
The $V_1$ amplitude from the 2D fit method is generally larger than the Fourier coefficient magnitude from the large-$\deta$ correlations.

The $V_2$ coefficient shown in Fig.~\ref{fig:Vn}(b) is positive and exhibits a decreasing trend with multiplicity. The trends may be suggestive of nonflow, which would be inversely proportional to multiplicity (e.g., nonflow contribution from the near-side peak). It is possible that the 2D fits are not adequate to separate the correlations into flow and nonflow in those \pp\ interactions, such that the fitted harmonics still reflect some nonflow characteristics. It is interesting to notice that the 2D fit $V_2$ values in the two lowest multiplicity bins of non-diffractive events are higher than the Fourier coefficients calculated at large $|\deta|>1.5$, whereas they are close to each other for the three higher multiplicity bins. This may reflect the difficulty in fitting the nonflow components in these low multiplicity events. The reason for this difficulty might be due to the fact that the near-side peak does not have a large presence and is barely noticeable for the two low multiplicity bins. 
It is also noteworthy that the fitted $V_1$ and $V_2$ parameters are anti-correlated, and this may play a role in the lowest multiplicity bin of the non-diffractive events. 
It may be at play for the diffractive events, where the three low multiplicity bins reveal larger fit $V_2$ and more negative fit $V_1$ than the Fourier counterparts. 

Finally, the $V_3$ coefficient shown in Fig.~\ref{fig:Vn}(c) is close to zero with large uncertainties. The overall magnitude may be negative, suggestive of influence of away-side peak. 
It is noteworthy that there seem no qualitative differences in the extracted $V_n$ harmonics between diffractive and non-diffractive events despite distinctive physics processes.

\begin{figure}[H]
	\centering
	\includegraphics[width=0.45\textwidth]{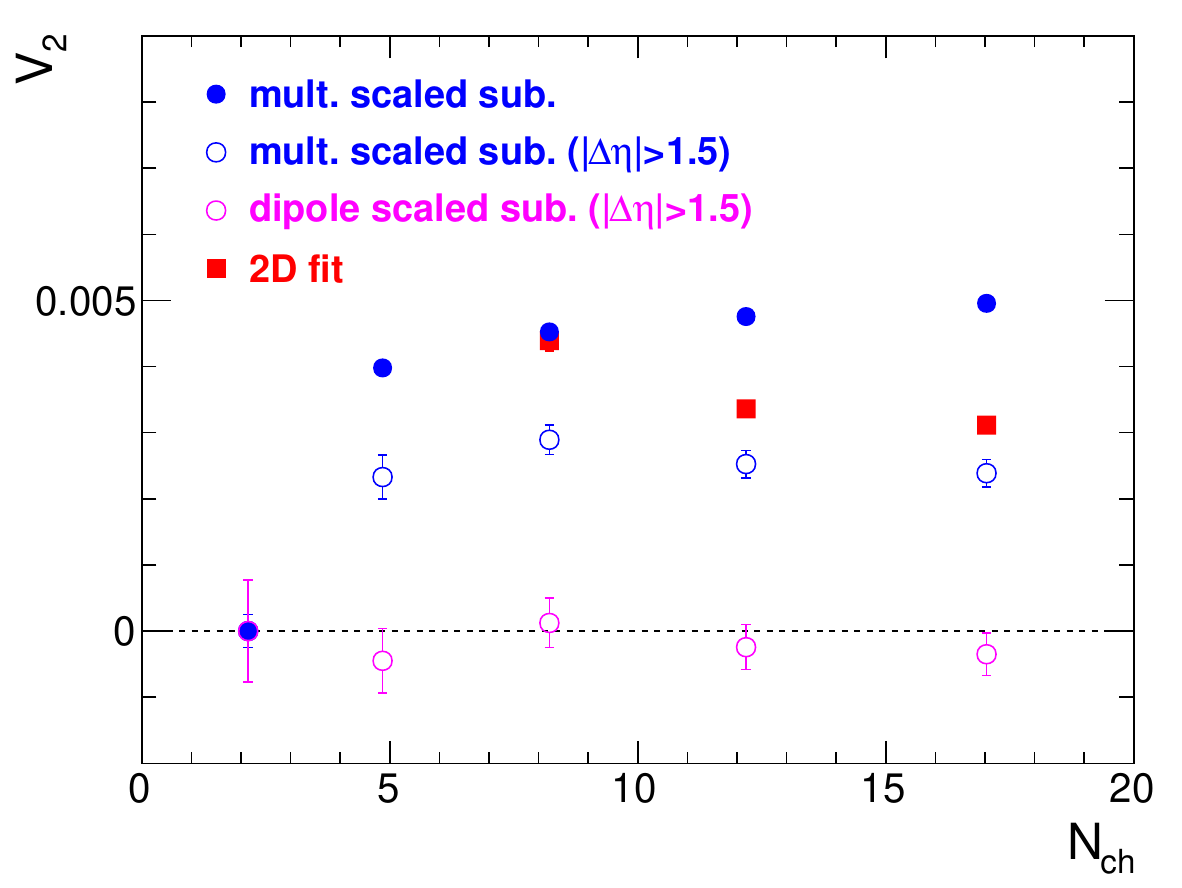}
	\caption{Low-multiplicity subtracted $V_2$ values as functions of multiplicity. Inverse multiplicity scaled subtraction (blue points) is shown for no $\deta$ cut and $|\deta|>1.5$. Dipole-scaled subtraction (pink points) is shown for $|\deta|>1.5$. Also shown are $V_2$ values from 2D fit (red squares). Results are obtained from SS correlation functions of non-diffractive events.}
	\label{fig:V2sub}
\end{figure}

One method to mitigate nonflow contributions is to subtract correlations in low-multiplicity events from those in high-multiplicity events~\cite{Feng:2024eos}. 
This is possible under the assumption that the nonflow correlation shape does not change with multiplicity and the abundance of the nonflow correlation sources is proportional to multiplicity~\cite{Feng:2024eos}. 
The latter also assumes, implicitly, that there are no biases to nonflow correlations from selection of events according to the event multiplicity. Under these assumptions, the ``flow'' harmonics in high-multiplicity events would be
\begin{equation}
V_n^{\rm sub} = V_n^{\rm high} - \frac{N^{\rm low}}{N^{\rm high}} V_n^{\rm low}\,,
\end{equation}
where the superscript `high' and `low' denote those quantities in high- and low-multiplicity events, respectively.
Figure~\ref{fig:V2sub} shows the $V_2^{\rm sub}$ as functions of multiplicity for non-diffractive events. The results are shown for two cases, one uses SS correlations without any $\deta$ cut, and the other uses those with $|\deta|>1.5$. The results differ, and the interpretation can be either (i) that they may reflect collective flow--which can be possible in \pp\ interactions because of multi-parton interactions or color reconnection--and there is a strong $\deta$ decorrelation effect, or (ii) that they may not reflect collective flow--which is naively expected to be $\deta$ independent due to its origin in early dynamics--but nonflow as nonflow is not independent of multiplicity or there are strong multiplicity biases in event selection. For comparison, the 2D fit $V_2$ values in the three high multiplicity bins are also plotted in Figure~\ref{fig:V2sub}; those for the lower multiplicity bins are significantly higher as seen from Fig.~\ref{fig:Vn}. 
No obvious systematics are apparent from the comparison.

The $V_2^{\rm sub}$ results shown in Fig.~\ref{fig:V2sub} from the various methods suggest the single-particle azimuthal anisotropy $v_2=\sqrt{V_2}$ of the order of 5--7\%. This is comparable to experimental findings at the LHC~\cite{CMS:2013jlh,Aad:2013fja,ALICE:2019zfl}, despite of the lower RHIC energy used in our simulation. This is especially the case for the $V_2^{\rm sub}$ results extracted from correlations in the large-$\deta$ region of $|\deta|>1.5$, the approximate region used in LHC experiments. It would be interesting to further investigate the possible anisotropic flows in Pythia simulations of \pp\ interactions at LHC energies~\cite{Bierlich:2018xfw}.

Another recipe for nonflow subtraction is to assume that the negative $V_1$ dipole is all nonflow and the nonflow correlation shape does not change with multiplicity, i.e., the nonflow $V_n$ ($n>1$) is proportional to $V_1$. This would lead to
\begin{equation}
V_n^{\rm sub} = V_n^{\rm high} - \frac{V_1^{\rm high}}{V_1^{\rm low}} V_n^{\rm low}\,.
\end{equation}
The $V_2^{\rm sub}$ values obtained from this dipole-scaled subtraction are shown by the pink points in Fig.~\ref{fig:V2sub} for SS correlations at $|\deta|>1.5$. 
They are all consistent with zero, implying that the $V_1$ and $V_2$ harmonics of the large-$\deta$ SS correlations scale with each other, which is indeed the case by comparing Fig.~\ref{fig:Vn}(a) and Fig.~\ref{fig:Vn}(b).

\section{Summary}
Azimuthal anisotropies have been observed in high-energy small-system (proton-proton and proton/deuteron/$^3$He-nucleus) collisions, similar to those measured in relativistic heavy ion collisions. This raises an intriguing possibility of collective flow behavior and the formation of a quark-gluon plasma droplet in those small-system collisions and has aroused intense research interest. Because of severe contamination of nonflow (genuine few-body) correlations in those small systems, analysis and interpretation of the two-particle azimuthal cumulant anisotropies are challenging. Several nonflow mitigation methods are available with various assumptions and degrees of validity. At least three interpretations are possible: (i) the measured anisotropies are part of nonflow correlations that differ from low- to high-multiplicity events; (ii) the measured anisotropies are a result of multi-parton interactions and color connections which are different from hydrodynamic-like interactions in the final state of heavy ion collisions; and (iii) there is an appreciable contribution from  hydrodynamic-like final-state interactions, thus collective flow, in those small-system collisions. 

To elucidate further, we studied two-particle angular correlations in pseudorapidity and azimuthal differences via simulated \pp\ interactions using the \pmodel\ event generator. The model includes multi-parton interactions and color reconnection, but no final-state interactions or hydrodynamic expansion. We present our study pedagogically and report azimuthal anisotropies extracted using several methods. These methods have different assumptions and sensitivities to nonflow correlations and, as such, our results represent a fair range in analysis method dependency and degree of validity. We observe finite azimuthal anisotropies from several, but not all, of the employed methods. We observe no qualitative difference between diffractive and non-diffractive events, despite their distinctive physics processes. The results are qualitatively, and in some cases semi-quantitatively, similar to those observed in experimental data. Our study, therefore, provides an important benchmark that can aid in improving data analysis and interpreting experimental findings.

Our study can be built up on and further enhanced in several ways. One is to turn off multi-parton interactions and color reconnection to examine the effects on the final azimuthal anisotropies. Another is to simulate \pp\ interactions at the higher LHC energies and expand to proton-nucleus and nucleus-nucleus collisions using the same Pythia model, and investigate the systematics of the azimuthal anisotropies on system size and collision energy. A thorough comparison of experimental data to Pythia simulations with variations in physics processes,  at the level of both the raw correlation functions and the extracted azimuthal anisotropies, should be performed. Some of such studies have already been reported,  but a systematic study would be invaluable.

\section*{Acknowledgment} 
MST acknowledges support from the Purdue-Colombia Exchange and Partnership Program. 
This work is supported in part by the U.S.~Department of Energy (Grant No.~DE-SC0012910).

\bibliography{ref}

\end{document}